\documentclass[12pt]{iopart}
\usepackage{etoolbox}
\patchcmd{\numparts}{\addtocounter{equation}{1}}{\refstepcounter{equation}}{}{}

\expandafter\let\csname equation*\endcsname\relax
\expandafter\let\csname endequation*\endcsname\relax
\usepackage{amsmath}
\usepackage{amssymb}
\usepackage[colorlinks=true,linkcolor=black, urlcolor = black, citecolor = black, anchorcolor = blue]{hyperref}
\usepackage{iopams}

\usepackage{amsfonts, amsthm, amssymb}

\usepackage{graphicx}
\usepackage{subcaption}
\usepackage{multirow}
\usepackage{xcolor}
\usepackage{xspace}
\usepackage{wasysym}
\usepackage[T1]{fontenc} % for the polish {\k{e}}
\newcommand{\ave}[1]{\left\langle{#1}\right\rangle}

\newcommand{\plaind}{\text{d}}

\newcommand{\Esref}[1]{Eqs.~(\ref{#1})}

\usepackage{appendix}

\usepackage{tikz}
\usetikzlibrary{decorations.pathmorphing}
\usetikzlibrary{decorations.markings}
\usetikzlibrary{arrows,shapes,snakes,automata,backgrounds,petri}
\usetikzlibrary{calc}
\usetikzlibrary{patterns}
\usetikzlibrary{decorations.text}
\tikzset{
xxtsubstrate/.style={decorate, 
line width=1pt,
draw=orange, 
decoration=snake, 
segment amplitude=0.75mm, 
line after snake=0.25mm,
line before snake=0.25mm
},
tsubstrate/.style={decorate, 
line width=1pt,
draw=orange, 
decoration=snake, 
segment amplitude=0.5mm, 
segment length=5pt,
segment amplitude=0.2mm, 
line after snake=1mm,
line before snake=1mm
},
Bsubstrate/.style={decorate, 
line width=1pt,
draw=orange, 
decoration=snake,
segment length=5pt,
segment aspect=0,
segment amplitude=0.5mm, 
line after snake=0mm,
line before snake=0mm
},
substrate/.style={decorate, 
line width=1pt,
draw=black, 
decoration=snake, 
segment length=5pt,
segment amplitude=0.5mm, 
line after snake=0.5mm,
line before snake=0.5mm
},
BCsubstrate/.style={decorate, 
line width=1pt,
draw=olive, 
decoration=snake,
segment length=5pt,
segment aspect=0,
segment amplitude=0.5mm, 
line after snake=0mm,
line before snake=0mm
},
Csubstrate/.style={decorate, 
line width=1pt,
draw=olive, 
decoration=snake, 
segment length=5pt,
segment amplitude=0.5mm, 
line after snake=0.5mm,
line before snake=0.5mm
},
activity/.style={very thick,draw=red,postaction={decorate},
decoration={markings,mark=at position .5 with
{\arrow[draw=red]{>}}}},
tactivity/.style={thick,draw=red,postaction={decorate},
decoration={markings,mark=at position .5 with
{\arrow[draw=red]{>}}}},
tEPSactivity/.style={thick,draw=red,postaction={decorate},
decoration={markings,mark=at position .55 with
{\arrow[draw=red]{>}}}},
tAactivity/.style={thick,draw=cyan},
Aactivity/.style={line width=1pt,draw=black},
Bactivity/.style={very thick,draw=blue,dashed},
Cactivity/.style={very thick,draw=red},
tSactivity/.style={thick,draw=cyan,postaction={decorate},
decoration={markings,mark=at position .7 with
{\arrow[draw=cyan]{>}}}},
Sactivity/.style={very thick,draw=cyan,postaction={decorate},
decoration={markings,mark=at position .7 with
{\arrow[draw=cyan]{>}}}},
polarity/.style={decorate, 
line width=1pt,
draw=red,
decoration={markings,mark=between positions 0 and 1 step 1.1mm with {\draw[red,thick]  (0,0) circle (0.03)} },
segment length=5pt,
segment amplitude=0.5mm, 
},
Bpolarity/.style={decorate, 
line width=1.5pt,
draw=red,
segment length=5pt,
segment amplitude=0.5mm, 
},
density/.style={ 
line width=1.5pt,
draw=black,
densely dashed,
segment length=5pt,
segment amplitude=0.5mm, 
}
}

\newcommand{\union}{\cup}
\newcommand{\tildephi}{\widetilde\phi}
\newcommand{\tildepsi}{\widetilde{\psi}}
\newcommand{\tilderho}{\widetilde{\rho}}
\newcommand{\tildenu}{\widetilde{\nu}}

\newcommand{\tildeu}{\widetilde{u}}

\newcommand\ptwiddle[1]{\mathord{\mathop{#1}\limits^{\scriptscriptstyle(\sim)}}}

\newcommand{\Cref}[1]{Chapter~\ref{#1}}

\newcommand{\cref}[1]{\ref{#1}}
\newcommand{\deltabar}{\delta\mkern-8mu\mathchar'26}
\newcommand{\imag}{\mathring{\imath}}
\newcommand{\dbar}{\plaind\mkern-6mu\mathchar'26}
\newcommand{\dint}[1]{\mathchoice{\!\plaind#1\,}{\!\plaind#1\,}{\!\plaind#1\,}{\!\plaind#1\,}}

\newcommand{\dintbar}[1]{\mathchoice{\!\dbar#1\,}{\!\dbar#1\,}{\!\dbar#1\,}{\!\dbar#1\,}}

\renewcommand{\exp}[1]{\mathchoice{e^{#1}}{\operatorname{exp}\!\left(#1\right)}{\operatorname{exp}\!\left(#1\right)}{\operatorname{exp}\!\left(#1\right)}}
\newcommand{\mA}{\mathcal{A}}
\newcommand{\mD}{\mathcal{D}}
\newcommand{\mF}{\mathcal{F}}
\newcommand{\mG}{\mathcal{G}}

\newcommand{\mL}{\mathcal{L}}
\newcommand{\mO}{\mathcal{O}}

\newcommand{\mW}{\mathcal{W}}

\newcommand{\entropyProduction}{\dot{S}_i}

\newcommand{\save}[1]{\langle {#1}\rangle}
\newcommand{\bave}[1]{\big\langle {#1}\big\rangle}
\newcommand{\Deff}{D_\text{eff}}

\newcommand{\dotxR}{\dot{x}_\phi}
\newcommand{\dotxB}{\dot{x}_\psi}
\newcommand{\Pphi}{P_\phi}
\newcommand{\Ppsi}{P_\psi}

\newcommand{\He}[2]{H_{#1}\left(#2\right)}

\usepackage{xspace}
\newcommand{\latin}[1]{{\it #1}}
\newcommand{\ie}{\latin{i.e.}\@\xspace}

\allowdisplaybreaks
\usepackage[square,numbers]{natbib}
\begin{document}

\title{Run-and-tumble motion: field theory and entropy production}

\author{Rosalba Garcia-Millan$^{1,2,3}$, Gunnar Pruessner$^{1,3}$}
 \ead{rg646@cam.ac.uk, g.pruessner@imperial.ac.uk}
\address{$^1$Department of Mathematics, Imperial College London, 180 Queen's Gate, London SW7 2AZ, UK.\\
$^2$DAMTP, Centre for Mathematical Sciences, University of Cambridge, Wilberforce Road, Cambridge CB3 0WA, UK.\\
$^3$Centre for Complexity Science, Imperial College London, UK.}

\begin{abstract}
Run-and-tumble motion is an example of active motility
where particles move at constant speed and change direction at random times.
In this work we study  run-and-tumble motion with 
diffusion in a harmonic potential in one dimension
via a path integral approach.
We derive a Doi-Peliti field theory and use it to calculate the entropy production and other observables 
in closed form.
All our results are exact.
\end{abstract}

\section{Introduction} 
Active matter encompasses reaction-diffusion systems out of equilibrium whose components are 
subject to local
non-thermal forces
\cite{FodorETAL:2016}.
There is a plethora of interesting patterns
and phenomenology in this broad class of systems.
Some fascinating
examples are active nematics \cite{MartinezIgnes-MullolCasademuntSagues:2019,StrubingETAL:2020,VafaETAL:2020},
active emulsions \cite{LeeWurtz:2018},
and active motility \cite{DauchotDemery:2019, BertrandETAL:2018},
amongst many others. 
Active matter has become a research focus in statistical mechanics
over recent years, 
as it addresses fundamental questions on the physics
of non-equilibrium systems,
but also as a quantitative approach to 
biological physics
and the path to designing an autonomous 
microbiological engine \cite{DiLeonardoETAL:2010,EkehCatesFodor:2020,PietzonkaETAL:2019,MarkovichETAL:2020}.

\begin{figure}
    \centering
    \includegraphics[width=0.49\textwidth]{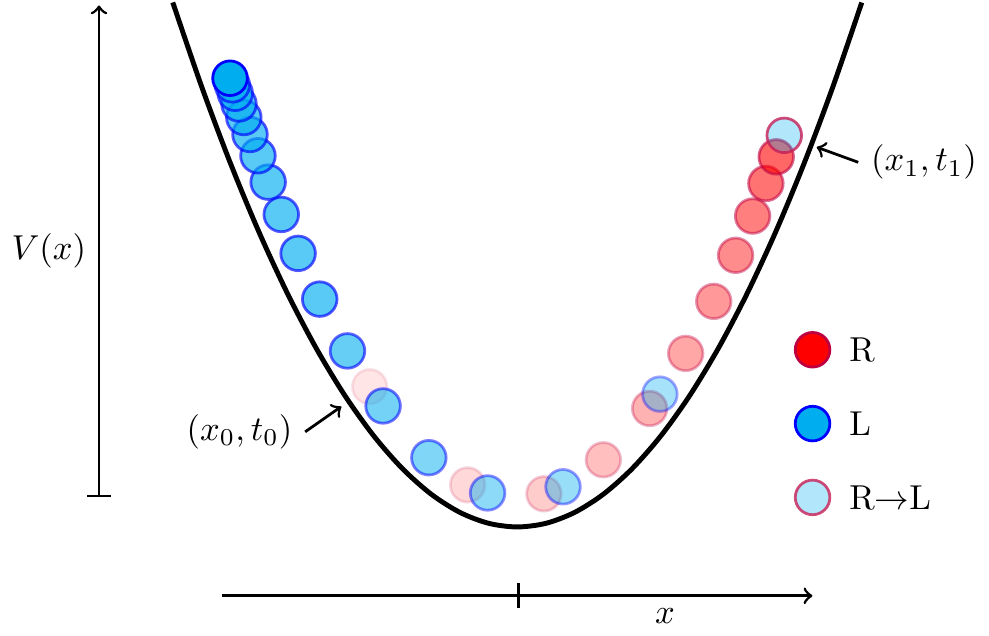}
    \caption{\textbf{Trajectory snapshots 
    of an RnT particle
    in a harmonic potential} $V(x)$ (solid line). Initially, the 
    particle  at $(x_0,t_0)$  is right-moving
    with positive 
    velocity $w$ (red circles), 
    until it tumbles at $(x_1,t_1)$ and
    becomes left-moving  with
    negative velocity $-w$ (blue circles).
    }
    \label{sketchRnT}
\end{figure}

In this paper we study a model of active motility known as \emph{run and tumble}
(RnT) that has been used to describe
bacterial swimming patterns  such as that of
\emph{Escherichia coli} and \emph{Salmonella}
 \cite{TailleurCates:2008,ElgetiWinklerGompper:2015}. 
A particle undergoing run-and-tumble motion 
 moves in a 
sequence of \emph{runs} at constant self-propulsion speed $w$ interrupted by sudden changes
(\emph{tumbles}) in its orientation 
that happen at Poissonian rate $\alpha$, 
\cite{SlowmanEvansBlythe:2017, RenadheerRoyGopalakrishnan:2019, Sevilla:2019, SolonCatesTailleur:2015, MallminBlytheEvans:2019, ShreshthaHarris:2019,BasuETAL:2020, LacroixMiron:2020}. 
This motion pattern is ballistic at the microscopic 
scale and diffusive at the large scale with effective 
diffusion constant $\Deff=w^2/\alpha$ \cite{Garcia-MillanPhD:2020,MalakarETAL:2018, CatesTailleur:2013, VillaETAL:2020, BasuETAL:2020, TailleurCates:2008,SantraBasuSabhapandit:2020,DeanMajumdarSchawe:2020}.
We study a run-and-tumble particle subject to thermal 
noise confined in a harmonic potential $V(x)=kx^2/2$,
see \fref{sketchRnT}.

We follow a path integral approach  
\cite{RenadheerRoyGopalakrishnan:2019}
whereby we  derive a perturbative field theory in the Doi-Peliti 
framework \cite{Taeuber:2014} and use it to calculate a number of observables in 
closed form, with an emphasis on the entropy production
\cite{Razin:2020}.
Despite our approach being perturbative, all our results are exact.
The presence of the external potential presents technical challenges in the
derivation of the field theory that
resemble those encountered in other contexts
such as the quantum harmonic oscillator \cite{Lindenbaum:1999,TaeuberHowardVollmayr-Lee:2005,WeberFrey:2017}. 
We use a combination of the Fourier transform and
Hermite polynomials to parametrise the fields as to
diagonalise the action functional.

We regard the Doi-Peliti framework as a method to solve a master equation,
or a Fokker-Planck equation.
It therefore crucially retains
the microscopic dynamics of the system and captures the particle entity.
For this reason, the Doi-Peliti framework provides a solid route to calculate the
entropy production and, most importantly, proves itself to be an effective
tool to study active particle systems.
In this paper, we illustrate this point
by studying RnT motion.

The contents of this paper are organised as follows: 
in Sec.~\ref{sec_rnt01}, we derive a field theory for an RnT particle in a 
harmonic potential; 
in Sec.~\ref{sec_entropy} we use this field theory to calculate the entropy
production; and in Sec.~\ref{sec_disc} we discuss our results.
In the appendix, we have included other relevant observables: 
mean square displacement (\ref{sec:app_MSD}),
two-point correlation functions (\ref{sec:app_corr_point} and \ref{sec:app_corr}),
 expected velocity (\ref{sec:app_vel}),
and stationary distribution (\ref{sec:app_distribution}).

\section{Field theory of run-and-tumble motion with diffusion in a harmonic potential}
\label{sec_rnt01} 
In one dimension, we can think of run-and-tumble motion as the 
interaction between right- and left-moving particles that 
transmute into one another at Poissonian rate $\alpha/2$,
see \fref{sketchRnT}.
The Langevin equations of each species
are
\begin{equation}\label{def_Langevin_phi}
 \dotxR= -\partial_x V(x)+w+\eta(t)   
\end{equation}
for right-moving particles 
and 
\begin{equation}\label{def_Langevin_psi}
 \dotxB= -\partial_x V(x)-w+\eta(t)   
\end{equation}
for left-moving particles, where $\eta$ is a Gaussian white noise with mean
$\ave{\eta(t)} =0$ and correlation function $\ave{\eta(t)\eta(t')}=2D\delta(t-t')$, 
with diffusion constant $D$.
The corresponding Fokker-Plank equations are coupled
due to the transmutation between species through a
gain and loss terms,
\begin{numparts}\label{FP_all}
\begin{eqnarray}
\label{rnt2FP}
\partial_t \Pphi= -\partial_x \left[ (-kx+w) \Pphi \right]  + D\partial_x^2 \Pphi +\frac{\alpha}{2}(\Ppsi - \Pphi)\,, 
\label{FPr}\\
\partial_t \Ppsi= -\partial_x \left[ (-kx-w) \Ppsi \right] + D\partial_x^2 \Ppsi +\frac{\alpha}{2}(\Pphi - \Ppsi)\,, 
\end{eqnarray}
\end{numparts}
where $\Pphi(x,t)$ and $\Ppsi(x,t)$ are the probability densities of  a right- and a left-moving particle
respectively, as a function of  position $x$ and time $t$. It follows that the action functional \cite{Garcia-MillanPruessner:2020} of this process is
\begin{eqnarray}
\mA
= \int \dint{x} \dint{t}
\big\{  \tildephi\left(\partial_t \phi +\partial_x \left[ (-kx+w) \phi \right]  - D\partial_x^2 \phi \right) \nonumber\\
\qquad +  \tildepsi\left(\partial_t \psi +\partial_x \left[ (-kx-w) \psi \right] - D\partial_x^2 \psi \right)
 +  \frac{\alpha}{2} (\tildephi - \tildepsi)(\phi-\psi) \big\} \,
\label{Arnt2phipsi}
\end{eqnarray}
where  
 $\phi$ is the annihilation field of right-moving particles;
 $\tildephi$ is the Doi-shifted \cite{Cardy:2008}  creation field $\phi^\dagger=\tildephi+1$ of right-moving particles;
 $\psi$ is the annihilation field of left-moving particles;
 and
 $\tildepsi$ is the Doi-shifted creation field $\psi^\dagger=\tildepsi+1$ of left-moving particles.
The action functional \eref{Arnt2phipsi} allows the  calculation of an observable $\bullet$ via the path integral
\begin{equation}
\ave{\bullet} = \int\mD[\phi,\tildephi,\psi,\tildepsi] \bullet\exp{-\mA([\phi,\tildephi,\psi,\tildepsi])  } \,.
\label{obser_T}
\end{equation}
The action in \Eref{Arnt2phipsi}  contains only bilinear terms, where the only interaction between 
species is due to transmutation. This action does not have any non-linear couplings.
Motivated by the matrix representation of the action
\eref{Arnt2phipsi}, we refer to any terms involving
$\tildephi\phi$ or $\tildepsi\psi$ as 
diagonal terms, and to the terms $\tildephi\psi$ and 
$\tildepsi\phi$ as off-diagonal terms.
The diagonal terms in \eref{Arnt2phipsi} are \emph{semi-local}
due to the derivatives in space and time, which need to be made
\emph{local} in order to carry out the Gaussian (path) integral.
This is usually achieved by expressing the
fields in Fourier space. In this case, however, Fourier-transforming 
\eref{Arnt2phipsi} yields the action local
in the frequency $\omega$ but not in reciprocate position $x$.
Due to the diagonal terms $-k\tildephi\partial_x(x\phi)$
$-k\tildepsi\partial_x(x\psi)$, the action remains semi-local 
after Fourier-transforming.

Instead, we first parametrise the action by the density field 
$\rho = (\phi+\psi)/\sqrt{2}$ and the polarity field
$\nu = (\phi-\psi)/\sqrt{2}$, also called 
chirality \cite{BijnensMaes:2020},
with the analogous transformation for the conjugate fields.
This change of variables is useful in other contexts and is known
under other names, such as the Keldysh rotation \cite{Kamenev:2011}. 
In our case, the advantage is that
the action remains invariant under the transformation
$\rho \leftrightarrow\nu$, $\tilderho \leftrightarrow\tildenu$, 
except for the mass term $\alpha\tildenu\nu$,
\begin{eqnarray}
\mA
= \int \dint{x} \dintbar{\omega} \big\{ 
-\imag\omega \tilderho \rho -k\tilderho\partial_x(x\rho)  -D\tilderho\partial_x^2\rho\nonumber\\ 
\qquad
 -\imag\omega\tildenu \nu  -k\tildenu\partial_x(x\nu) -D\tildenu\partial_x^2\nu 
+\alpha\tildenu \nu
  + w \tilderho\partial_x\nu + w \tildenu\partial_x\rho \big\} \,.
\label{action_rhonu_nonlocal}
\end{eqnarray}
We then use the solution to the eigenvalue problem 
$\mL[u(x)] = \lambda u(x)$, with 
\begin{equation}\label{def_Lop}
    \mL[u(x)] = \partial_x^2 u(x)+\partial_x(x\,u(x))/L^2 \ ,
\end{equation}  $L=\sqrt{D/k}$ and $\lambda\in\mathbb{R}$,
where the differential operator follows from the diagonal terms
in \eref{action_rhonu_nonlocal}.
By using the fields $\rho$ and $\nu$,
instead of the original $\phi$ and $\psi$, we have the same differential
operator $\mL$ acting on both 
$\rho$ and $\nu$.

The solution to this eigenvalue problem is 
the set of Hermite functions
\begin{equation}
u_n(x) = \exp{-\frac{x^2}{2L^2}} \, \He{n}{\frac{x}{L}} \,,
\end{equation}
with eigenvalue $\lambda_n = -{n}/{L^2}$, 
where $\He{n}{x}$ is the $n$-th Hermite polynomial, 
see  
\ref{sec:sec_Hermite} \cite{WalterPruessnerSalbreux:2020,RiskenFrank:1996,Lindenbaum:1999}.
Defining the set of functions
\begin{equation}
\tildeu_n(x) = \frac{1}{\sqrt{{2\pi}} \, n! }{\He{n}{\frac{x}{L}}} \,,
\end{equation}
the orthogonality relation between
$u_n(x)$ and $\tildeu_m(x)$ follows from
the orthogonality 
 of Hermite polynomials in \Eref{He_orth}, 
\begin{equation}
\label{uv_orth}
\int_{-\infty}^\infty\dint{x} u_n(x) \tildeu_m(x) = L\, \delta_{n,m}\,,
\end{equation}
where $\delta_{n,m}$ is the Kronecker $\delta$.
We can now use  $u_n$ and $\tildeu_m$ as  basis for the fields
 $\rho$, $\tilderho$, $\nu$ and $\tildenu$,
\begin{numparts}\label{parametrisation_all}
\begin{eqnarray}
\label{parametrisation}
\ptwiddle{\rho}(x,t) =  \int\dintbar{\omega} \exp{-\imag\omega t} \frac{1}{L}\sum_n \ptwiddle{\rho}_n(\omega)\ptwiddle{u}_n(x) \,, 
\label{decomp_alpha}\\
\ptwiddle{\nu}(x,t) =  \int\dintbar{\omega} \exp{-\imag\omega t} \frac{1}{L}\sum_n \ptwiddle{\nu}_n(\omega)\ptwiddle{u}_n(x) \,.
\end{eqnarray}
\end{numparts}

Using the representation \eref{parametrisation_all} and the 
orthogonality relation \eref{uv_orth} in the action $\mA = \mA_0+\mA_1$ in
\eref{action_rhonu_nonlocal}, 
we have
\begin{numparts}\label{mArhonu_all}
\begin{eqnarray}
\label{mArhonu}
\fl
\mA_0  = 
\frac{1}{L^2}\sum_{n,m} L\delta_{n,m}
 \int\dintbar{\omega}\dintbar{\omega'} \deltabar(\omega+\omega') \nonumber \\\fl\qquad
\times
\big[
\left(-\imag\omega + kn \right) \tilde{\rho}_n(\omega'){\rho}_m(\omega) +
\left(-\imag\omega + kn  + \alpha\right) \tilde{\nu}_n(\omega'){\nu}_m(\omega)\big] \,,
\label{action_A0}\\\fl
\mA_1  = 
- \frac{w}{L}
\frac{1}{L^2}\sum_{n,m}  L\delta_{n-1,m}  \int\dintbar{\omega}\dintbar{\omega'} \deltabar(\omega+\omega') 
\big[\tilde{\rho}_n(\omega'){\nu}_m(\omega)+ \tilde{\nu}_n(\omega'){\rho}_m(\omega) \big] \,,
\label{mA1rhonu}
\end{eqnarray}
\end{numparts}
where $\mA_0$ contains the local, diagonal terms
and $\mA_1$ contains the off-diagonal terms, which
are non-local due to $\delta_{n-1,m}$.
We can then regard $\mA_0$ as the Gaussian model and $\mA_1$ 
as a perturbation about it.

The Gaussian model corresponds to 
Ornstein-Uhlenbeck particles where one species has
decay rate $\alpha$.
Defining an observable $\bullet$ in
the Gaussian model as
\begin{equation}
    \ave{\bullet}_0
    = \int\mD[\rho,\tilderho,\nu,\tildenu] \bullet
    \exp{-\mA_0([\rho,\tilderho,\nu,\tildenu])} \,,
    \label{Gaussian_model}
\end{equation}
from \eref{obser_T} it follows that 
the perturbation expansion of the
observable in the full model, the RnT particle,
is
\begin{equation}
\label{obs_exp}
    \ave{\bullet} = 
    \ave{\bullet \exp{-\mA_1}}_0
    = \sum_{N=0}^\infty
    \frac{1}{N!}
    \ave{\bullet (-\mA_1)^N}_0 \,.
\end{equation}

By performing the Gaussian path integral \cite{LeBellac:1991} in \eref{Gaussian_model},
the bare propagators  read  
 \begin{numparts}\label{propaabb0_all}
 \begin{eqnarray}
  \label{propaabb0}
\tikz[baseline=-2.5pt]{
\node at (0.75,0) [above] {$\tilde{\rho}_m(\omega')$};
\node at (-0.75,0) [above] {${\rho}_n(\omega)$};
\draw[Aactivity] (0.75,0) -- (-0.75,0);
}\,
\hat{=} \save{\rho_n(\omega)\tilde{\rho}_m(\omega')}_0 
= \frac{L\delta_{n,m}\deltabar(\omega+\omega')}{-\imag \omega  +  kn  + r}  \,, \label{propaa}\\
\tikz[baseline=-2.5pt]{
\node at (0.75,0) [above] {$\tilde{\nu}_m(\omega')$};
\node at (-0.75,0) [above] {${\nu}_n(\omega)$};
\draw[substrate] (0.75,0) -- (-0.75,0);
}\,
\hat{=} \save{\nu_n(\omega)\tilde{\nu}_m(\omega')}_0 
= \frac{L\delta_{n,m}\deltabar(\omega+\omega')}{-\imag \omega  +  kn  + \alpha} \,,
 \end{eqnarray}
\end{numparts}
where $r$ is a mass term added to regularise the infrared divergence and which is to be taken to $0$
when calculating any observable. 
We use Feynman diagrams to represent propagators \cite{LeBellac:1991,Cardy:2008,Taeuber:2014},
where time (causality) is read from right to left.
On the other hand, we have
\begin{numparts}\label{propabba0_all}
\begin{eqnarray}
 \label{propabba0}
 \save{\rho_n(\omega)\tilde{\nu}_m(\omega')}_0 
=  \, 0  \,, \label{propab_0}\\
 \save{\nu_n(\omega)\tilde{\rho}_m(\omega')}_0 
=  \, 0  \,,\label{propba_0}
\end{eqnarray}
\end{numparts}
for any $n,m$, which implies that there is no
"interaction" between $\rho$ and $\nu$ at the bare
level. The perturbative part $\mA_1$ of the action, \Eref{mA1rhonu}, however, provides the amputated vertices
\begin{equation}\label{perturbative_vertex}
    \tikz[baseline=-2.5pt]{
\node at (-0.65,0) [above] {$\tilde{\rho}_n(\omega')$};
\node at (0.65,0) [above] {${\nu}_m(\omega)$};
\draw[substrate] (0.5,0) -- (0,0);
\draw[Aactivity] (0,0) -- (-0.5,0);
}
\hat{=} 
-\frac{w}{L} L \delta_{n-1,m}\deltabar(\omega+\omega')
\hat{=}
    \tikz[baseline=-2.5pt]{
\node at (-0.65,0) [above] {$\tilde{\nu}_n(\omega')$};
\node at (0.65,0) [above] {${\rho}_m(\omega)$};
\draw[Aactivity] (0.5,0) -- (0,0);
\draw[substrate] (0,0) -- (-0.5,0);
} \ ,
\end{equation}
which shift the index by one.

\subsection{Full propagator in reciprocate space}
To calculate certain observables such as the time-dependent 
probability distribution of the RnT particle,
we need the full propagators.
To derive the full propagators we use
the action in \eref{mArhonu_all}, the bare propagators in 
\eref{propaabb0_all}, \eref{propabba0_all}, the perturbative vertex \eref{perturbative_vertex}, as well as Wick's Theorem \cite{LeBellac:1991}.
Consider, for instance, the propagator of the density field
$\ave{\rho\tilderho}$. From \eref{obs_exp} we have
\begin{equation}
\label{propa_expansion}
\tikz[baseline=-2.5pt, scale=1]{
\draw[Aactivity] (0.75,0) -- (-0.75,0) node [at start, above] {$_m,\omega'$} node [at end, above] {$_n,\omega$};
\draw[black,fill=black] (0,0) circle (0.1);
} 
\hat{=}
\ave{\rho_n(\omega)\tilde{\rho}_m(\omega')}
=\sum_{N=0}^\infty \frac{1}{N!}
\ave{\rho_n(\omega)\tilde{\rho}_m(\omega')({-\mA_1})^N}_0 \,,
\end{equation}
where the zeroth order term 
is given in \Eref{propaa} and the first order term is
\begin{equation}
\ave{\rho_n(\omega)\tilde{\rho}_m(\omega')(-\mA_1)}_0 
 = \, 0 \,.
\end{equation}
\Eref{propa_expansion} allows us to calculate the stationary distribution, \ref{sec:app_distribution}.
The second order term in \eref{propa_expansion} is
\begin{eqnarray}
\tikz[baseline=-2.5pt, scale=1]{
\draw[Aactivity] (0.75,0) -- (0.25,0) node [at start, above] {$_m,\omega'$} ;
\draw[substrate] (0.25,0) -- (-0.25,0)  ;
\draw[Aactivity] (-0.25,0) -- (-0.75,0) node [at end, above] {$_n,\omega$};
}
\hat{=}
\frac{1}{2}\ave{\rho_n(\omega)\tilde{\rho}_m(\omega')(-\mA_1)^2}_0 \nonumber\\
 =  \frac{w^2}{L}  
\frac{\delta_{n,m+2}\deltabar(\omega+\omega')}{\left(-\imag \omega  +  k(m+2)  + r\right)\left(-\imag \omega  +  k(m+1)  +  \alpha\right)\left(-\imag \omega  +  km  + r\right)} \,,
\end{eqnarray}
using two of the index-shifting vertices \eref{perturbative_vertex}.
From \eref{propabba0} it follows 
that any  term  in \eref{propa_expansion} of odd order $N$ vanishes because the fields $\rho$, $\tilde{\rho}$,
$\nu$ and $\tilde{\nu}$ cannot be paired according to
\eref{propaabb0_all}. Then, the contributions to the full propagator
$\ave{\rho\tilderho}$ are
\begin{equation}
\frac{1}{N!}\ave{\rho_n(\omega)\tilderho_m(\omega')(-\mA_1)^N}_0 
= \left\{
\begin{array}{l l}
L\delta_{n,m+N}\deltabar(\omega+\omega')
\left(\frac{w}{L}\right)^N 
\prod_{j=0}^N \frac{1}{-\imag\omega + k(m+j) + p_j}
& \text{if } N \text{ even} \,,\\
0     & \text{if } N \text{ odd}\,, 
\end{array}
\right.
\label{prop_rhorhoN}
\end{equation}
where $p_j=r$ if $j$ is even and $p_j=\alpha$ if $j$ is odd.
Similarly, the contributions to the full propagators
$\ave{\nu\tildenu}$, $\ave{\rho\tildenu}$ and $\ave{\nu\tilderho}$
are, respectively,
\numparts
\begin{eqnarray}
\fl
\frac{1}{N!}\ave{\nu_n(\omega)\tildenu_m(\omega')(-\mA_1)^N}_0 
= \left\{
\begin{array}{l l}
L\delta_{n,m+N}\deltabar(\omega+\omega')
\left(\frac{w}{L}\right)^N 
\prod_{j=0}^N \frac{1}{-\imag\omega + k(m+j) + q_j}
& \text{if } N \text{ even} \,,\\
0     & \text{if } N \text{ odd}\,, 
\end{array}
\right.
\\
\fl
\frac{1}{N!}\ave{\rho_n(\omega)\tildenu_m(\omega')(-\mA_1)^N}_0 \label{prop_rhonuN}
= \left\{
\begin{array}{l l}
0 & \text{if } N \text{ even} \,,\\
L\delta_{n,m+N}\deltabar(\omega+\omega')
\left(\frac{w}{L}\right)^N 
\prod_{j=0}^N \frac{1}{-\imag\omega + k(m+j) + q_j}
& \text{if } N \text{ odd} \,,
\end{array}
\right.
\\
\fl
\frac{1}{N!}\ave{\nu_n(\omega)\tilderho_m(\omega')(-\mA_1)^N}_0 
= \left\{
\begin{array}{l l}
0 & \text{if } N \text{ even}\,, \\
L\delta_{n,m+N}\deltabar(\omega+\omega')
\left(\frac{w}{L}\right)^N 
\prod_{j=0}^N \frac{1}{-\imag\omega + k(m+j) + p_j}
& \text{if } N \text{ odd} \,,
\end{array}
\right.
\label{prop_nurhoN}
\end{eqnarray}
\endnumparts
where $q_j=\alpha$ if $j$ is even and $q_j=r$ if $j$ is odd.
The diagrammatic representation of the full propagators is
\begin{numparts}
\label{prop_expansion_all}
\begin{eqnarray}
\label{propaa_expansion}
\tikz[baseline=-2.5pt, scale=1]{
\draw[Aactivity] (0.75,0) -- (-0.75,0);
\draw[black,fill=black] (0,0) circle (0.1);
}
= 
\tikz[baseline=-2.5pt, scale=1]{
\draw[Aactivity] (0.75,0) -- (-0.75,0);
}
+
\tikz[baseline=-2.5pt, scale=1]{
\draw[Aactivity] (0.75,0) -- (0.25,0);
\draw[substrate] (0.25,0) -- (-0.25,0)  ;
\draw[Aactivity] (-0.25,0) -- (-0.75,0);
}
+
\tikz[baseline=-2.5pt, scale=1]{
\draw[Aactivity] (0.75,0) -- (0.45,0);
\draw[substrate] (0.45,0) -- (0.15,0)  ;
\draw[Aactivity] (0.15,0) -- (-0.15,0)  ;
\draw[substrate] (-0.15,0) -- (-0.45,0)  ;
\draw[Aactivity] (-0.45,0) -- (-0.75,0);
}
+\ldots \,,
\\
\tikz[baseline=-2.5pt, scale=1]{
\draw[substrate] (0.75,0) -- (-0.75,0);
\draw[black,fill=black] (0,0) circle (0.1);
}
= 
\tikz[baseline=-2.5pt, scale=1]{
\draw[substrate] (0.75,0) -- (-0.75,0);
}
+
\tikz[baseline=-2.5pt, scale=1]{
\draw[substrate] (0.75,0) -- (0.15,0);
\draw[Aactivity] (0.15,0) -- (-0.15,0)  ;
\draw[substrate] (-0.15,0) -- (-0.75,0);
}
+
\tikz[baseline=-2.5pt, scale=1]{
\draw[substrate] (0.75,0) -- (0.45,0);
\draw[Aactivity] (0.45,0) -- (0.15,0)  ;
\draw[substrate] (0.15,0) -- (-0.15,0)  ;
\draw[Aactivity] (-0.15,0) -- (-0.45,0)  ;
\draw[substrate] (-0.45,0) -- (-0.75,0);
}
+\ldots \,,
\\
\label{propab_expansion}
\tikz[baseline=-2.5pt, scale=1]{
\draw[substrate] (0.75,0) -- (0,0);
\draw[Aactivity] (0,0) -- (-0.75,0);
\draw[black,fill=black] (0,0) circle (0.1);
}
=
\tikz[baseline=-2.5pt, scale=1]{
\draw[substrate] (0.75,0) -- (0,0);
\draw[Aactivity] (0,0) -- (-0.75,0);
}
+
\tikz[baseline=-2.5pt, scale=1]{
\draw[substrate] (0.75,0) -- (0.37,0);
\draw[Aactivity] (0.37,0) -- (0,0) ;
\draw[substrate] (0,0) -- (-0.38,0) ;
\draw[Aactivity] (-0.38,0) -- (-0.75,0);
}
+
\tikz[baseline=-2.5pt, scale=1]{
\draw[substrate] (0.75,0) -- (0.45,0);
\draw[Aactivity] (0.45,0) -- (0.25,0)  ;
\draw[substrate] (0.25,0) -- (-0.05,0)  ;
\draw[Aactivity] (-0.05,0) -- (-0.25,0)  ;
\draw[substrate] (-0.25,0) -- (-0.55,0) ;
\draw[Aactivity] (-0.55,0) -- (-0.75,0);
}
+\ldots \,,
\\
\tikz[baseline=-2.5pt, scale=1]{
\draw[Aactivity] (0.75,0) -- (0,0);
\draw[substrate] (0,0) -- (-0.75,0);
\draw[black,fill=black] (0,0) circle (0.1);
}
=
\tikz[baseline=-2.5pt, scale=1]{
\draw[Aactivity] (0.75,0) -- (0,0);
\draw[substrate] (0,0) -- (-0.75,0);
}
+
\tikz[baseline=-2.5pt, scale=1]{
\draw[Aactivity] (0.75,0) -- (0.38,0);
\draw[substrate] (0.38,0) -- (0,0) ;
\draw[Aactivity] (0,0) -- (-0.37,0) ;
\draw[substrate] (-0.37,0) -- (-0.75,0);
}
+
\tikz[baseline=-2.5pt, scale=1]{
\draw[Aactivity] (0.75,0) -- (0.55,0);
\draw[substrate] (0.55,0) -- (0.25,0)  ;
\draw[Aactivity] (0.25,0) -- (0.05,0)  ;
\draw[substrate] (0.05,0) -- (-0.25,0)  ;
\draw[Aactivity] (-0.25,0) -- (-0.45,0) ;
\draw[substrate] (-0.45,0) -- (-0.75,0);
}
+\ldots \,,
\label{propba_expansion}
\end{eqnarray}
\end{numparts}
where the black circle 
\tikz[baseline=-2.5pt, scale=1]{
\draw[black,fill=black] (0,0) circle (0.1);
}
represents the sum over all possible diagrams that have the 
same incoming and outgoing legs.
Some of these propagators are calculated in closed form in \ref{sec_propclosedform}.

\newcommand{\somepropagatorsclosedform}{\Esref{prop_aan}--\eref{prop_babn}}

\subsection{Short-time propagator in real space}
In this section we calculate the
short-time propagator $\save{\phi(y,\tau)\tildephi(x,0)}_1$
of a 
right-moving particle that moves from 
position $x$ to $y$ in an interval of 
time $\tau$,
and the
short-time propagator $\save{\psi(y,\tau)\tildephi(x,0)}_1$
of a 
right-moving particle $x$ that transmutates
into a left-moving particle at $y$
 in an interval of 
time $\tau$.
The subindex indicates that the propagator is 
expanded to first order about the Gaussian model,
$\save{\bullet}_1=\save{\bullet}_0-\save{\bullet\mA_1}_0$. Expanding to $N$th order in the perturbative part of the action generally provides the $N$th order in $\tau$ \cite{Garcia-MillanPruessner:2020}.
Using the recurrence relation \eref{recursive},
Mehler's formula \eref{Mehler} and
the propagators in \somepropagatorsclosedform,
these two propagators are,
\begin{numparts}
\label{short_props}
\begin{eqnarray}
\fl
     \frac{1}{2} 
    (\tikz[baseline=-2.5pt, scale=0.75]{
\draw[Aactivity] (0.75,0) -- (-0.75,0) ;
}+
\tikz[baseline=-2.5pt, scale=0.75]{
\draw[substrate] (0.75,0) -- (-0.75,0) ;
}+
\tikz[baseline=-2.5pt, scale=0.75]{
\draw[Aactivity] (0,0) -- (-0.75,0) ;
\draw[substrate] (0.75,0) -- (0,0) ;
}+\tikz[baseline=-2.5pt, scale=0.75]{
\draw[Aactivity] (0.75,0) -- (0,0) ;
\draw[substrate] (0,0) -- (-0.75,0) ;
}) 
    \hat{=} 
\save{\phi(y,\tau)\tildephi(x,0)}_1
\\
\fl
    =  \frac{1}{2}
    \sqrt{\frac{k}{2\pi D\left(1-\exp{-2k\tau}\right)}}
    \exp{-\frac{k(y-x\exp{-k\tau})^2}{2D\left(1-\exp{-2k\tau}\right)}}
    \nonumber\\
\fl
    \times
    \left(
    1+\exp{-\alpha\tau}
    + \frac{w k \left(y-x e^{-k\tau}\right)}{D\left(1- e^{-2k\tau}\right)}  
    \left[\frac{1}{k-\alpha}\left(
    \exp{-\alpha\tau} -\exp{-k\tau}\right)
    +\frac{1}{k+\alpha}\left(1 -\exp{-(k+\alpha)\tau}
    \right)
    \right]
    \right)\label{R_shortTimeprop_exact}\\
\fl
    = 
    \frac{1}{\sqrt{4\pi D\tau}}
    \exp{-\frac{(y-x)^2}{4D\tau}}
    \left(
    1+\frac{w(y-x)}{2D}
    + \left(-\frac{\alpha}{2} +\frac{w}{4D}
    \left(k(x+y)-\alpha(y-x)\right)\right)\tau
    +\mO\left( \tau^2 \right)
    \right)\,,
    \label{R_shortTimeprop}\\
\fl
     \frac{1}{2} 
    (\tikz[baseline=-2.5pt, scale=0.75]{
\draw[Aactivity] (0.75,0) -- (-0.75,0) ;
}-
\tikz[baseline=-2.5pt, scale=0.75]{
\draw[substrate] (0.75,0) -- (-0.75,0) ;
}+
\tikz[baseline=-2.5pt, scale=0.75]{
\draw[Aactivity] (0,0) -- (-0.75,0) ;
\draw[substrate] (0.75,0) -- (0,0) ;
}-
\tikz[baseline=-2.5pt, scale=0.75]{
\draw[Aactivity] (0.75,0) -- (0,0) ;
\draw[substrate] (0,0) -- (-0.75,0) ;
}) \hat{=}  \save{\psi(y,\tau)\tildephi(x,0)}_1
    \\
\fl
    =  \frac{1}{2}
    \sqrt{\frac{k}{2\pi D\left(1-\exp{-2k\tau}\right)}}
    \exp{-\frac{k(y-x\exp{-k\tau})^2}{2D\left(1-\exp{-2k\tau}\right)}}
    \nonumber\\
\fl
    \times
    \left(
    1-\exp{-\alpha\tau}
    + \frac{w k \left(y-x e^{-k\tau}\right)}{D\left(1- e^{-2k\tau}\right)}  
    \left[\frac{1}{k-\alpha}\left(
    \exp{-\alpha\tau} -\exp{-k\tau}\right)
    -\frac{1}{k+\alpha}\left(1 -\exp{-(k+\alpha)\tau}
    \right)
    \right]
    \right)\\
\fl
    =
    \frac{1}{\sqrt{4\pi D\tau}}
    \exp{-\frac{(y-x)^2}{4D\tau}}
    \left(\frac{\alpha}{2}\tau
    +\mO\left( \tau^2 \right)
    \right) \,.
    \label{S_shortTimeprop}
\end{eqnarray}
\end{numparts}
Comparing \eref{R_shortTimeprop} with \eref{S_shortTimeprop}
we see that the transition probability that involves exactly one
transmutation event, independently of displacement, is 
of higher order in $\tau$ than the transition probability of 
just a displacement.

\subsection{Zeroth, first and second moments of the 
position of a right-moving particle} 
\label{sec:app_momsposition}

\begin{figure}
    \centering
    \includegraphics[width=0.6\textwidth]{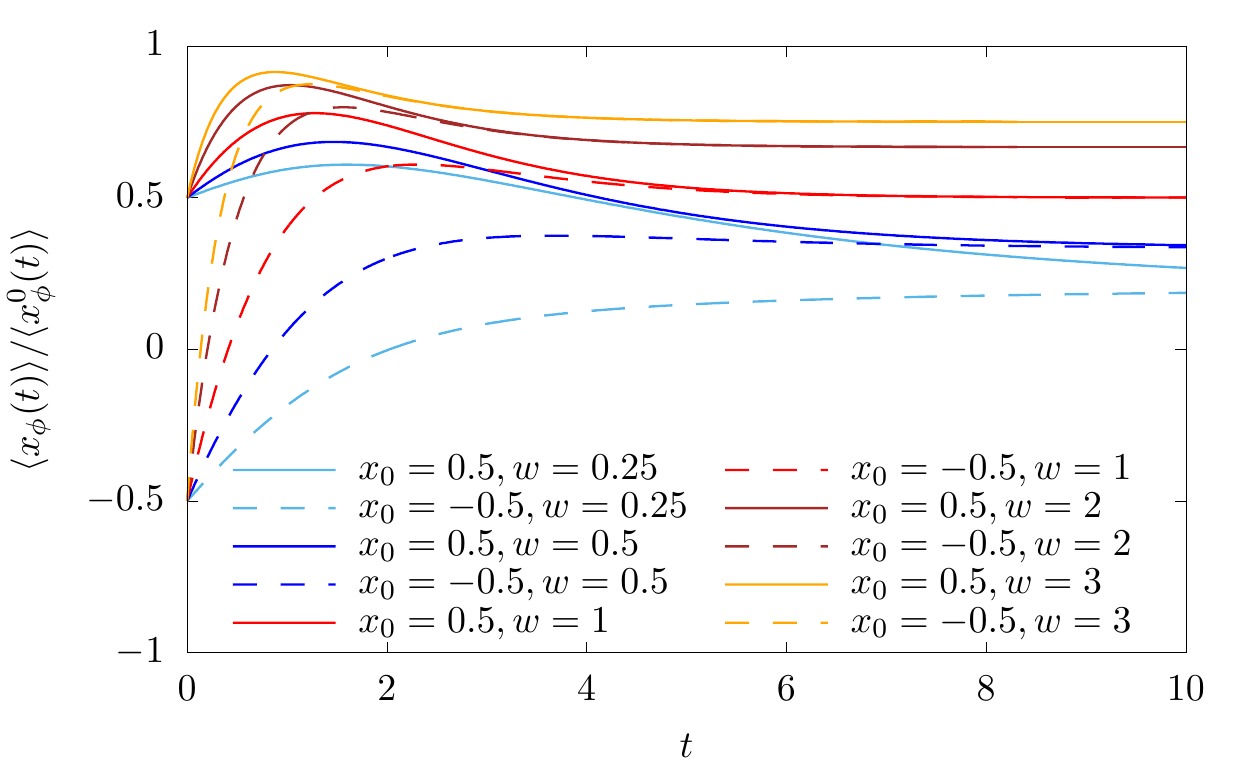}
    \caption{ \textbf{Conditional expected position $\ave{x_\phi(t)}/\ave{x_\phi^0(t)}$ of a right-moving RnT particle in a harmonic potential}, \Esref{PRt} and \eqref{xRt}, for a range of 
    $w$ and $x_0$, with  $\alpha=1$ and $k=w/\xi$     with $\xi=1$. }
    \label{fig_firstmomphi}
\end{figure}

As it will become clear in Sec.~\ref{sec_entropy}, 
we need the zeroth, first and second moments of the position
of a right-moving particle to calculate the entropy production
of an RnT particle in a harmonic potential. 
Assuming that the system is initialised 
with a right-moving particle placed at $x_0$ at time $t_0=0$,
the $n$th moment of its position is
\begin{equation}\label{nth_moment}
    \ave{x^n_\phi(t)} =  \int_{-\infty}^\infty
    \dint{x} x^n\save{\phi(x,t)\tildephi(x_0,0)}\,,
\end{equation}
where the propagator $\save{\phi(x,t)\tildephi(x_0,0)}$
expressed in terms of the $\ptwiddle{\rho}$ and $\ptwiddle{\nu}$ fields
contains the four full propagators,
$\save{\phi\tildephi} = (
\tikz[baseline=-2.5pt, scale=0.5]{
\draw[Aactivity] (0.75,0) -- (-0.75,0) ;
\draw[black,fill=black] (0,0) circle (0.1);
}+
\tikz[baseline=-2.5pt, scale=0.5]{
\draw[substrate] (0.75,0) -- (-0.75,0) ;
\draw[black,fill=black] (0,0) circle (0.1);
}+
\tikz[baseline=-2.5pt, scale=0.5]{
\draw[Aactivity] (0,0) -- (-0.75,0) ;
\draw[substrate] (0.75,0) -- (0,0) ;
\draw[black,fill=black] (0,0) circle (0.1);
}+\tikz[baseline=-2.5pt, scale=0.5]{
\draw[Aactivity] (0.75,0) -- (0,0) ;
\draw[substrate] (0,0) -- (-0.75,0) ;
\draw[black,fill=black] (0,0) circle (0.1);
})/2$, \Eref{prop_expansion_all}.
Definition \eref{nth_moment} contains a mild abuse of notation that we will keep committing throughout, as the angular brackets were introduced in \eqref{obser_T} as a path integral but are expectation over a density in \eqref{nth_moment}.
Since there is an integral over space and $x^n$
is a polynomial in $x$, the observable can be written  
as linear combinations of Hermite polynomials, which simplifies
the calculations by virtue of the orthogonality relation in
\eref{He_orth}. In particular, for the first three moments, 
$1=\He{0}{x/L}$, 
$x/L=\He{1}{x/L}$ and 
$(x/L)^2=\He{0}{x/L} + \He{2}{x/L}$.
Using the representation in \eref{parametrisation_all} and the
propagators derived in \ref{sec_propclosedform}, the
zeroth, first (see \fref{fig_firstmomphi}) and second moments are 
\begin{numparts}\label{all_moments}
\begin{eqnarray}
\fl
\ave{x_\phi^0(t)} 
=  \frac{\tildeu_0(x_0)}{2L^2}
\int_{-\infty}^\infty \dint{x} \He{0}{\frac{x}{L}}u_0(x) 
\left(
    \tikz[baseline=-2.5pt,scale=0.75]{
\draw[Aactivity] (0.75,0) -- (-0.75,0) node [at start, above] {$_0,0$} node [at end, above] {$_0,t$};
} +
    \tikz[baseline=-2.5pt,scale=0.75]{
\draw[substrate] (0.75,0) -- (-0.75,0) node [at start, above] {$_0,0$} node [at end, above] {$_0,t$};
}
    \right)    \nonumber\\
    =\frac{1}{2}\left(1+\exp{-\alpha t}\right)
    \,, 
    \label{PRt} \\
\fl
  \ave{x_\phi(t)} = 
\frac{1}{2L}\int_{-\infty}^\infty \dint{x} \He{1}{\frac{x}{L}}u_1(x) 
\nonumber\\\times
\left[
\left(
    \tikz[baseline=-2.5pt,scale=0.75]{
\draw[Aactivity] (0.75,0) -- (-0.75,0) node [at start, above] {$_1,0$} node [at end, above] {$_1,t$};
} +
    \tikz[baseline=-2.5pt,scale=0.75]{
\draw[substrate] (0.75,0) -- (-0.75,0) node [at start, above] {$_1,0$} node [at end, above] {$_1,t$};
}
    \right) \tildeu_1(x_0) 
    +
    \left(
    \tikz[baseline=-2.5pt,scale=0.75]{
\draw[Aactivity] (0,0) -- (-0.75,0) node [at end, above] {$_{1},t$};
\draw[substrate] (0.75,0) -- (0,0) node [at start, above] {$_0,0$};
} +
\tikz[baseline=-2.5pt,scale=0.75]{
\draw[substrate] (0,0) -- (-0.75,0) node [at end, above] {$_{1},t$};
\draw[Aactivity] (0.75,0) -- (0,0) node [at start, above] {$_0,0$};
}
\right) \tildeu_0(x_0)\right]    \nonumber \\
=
    \frac{x_0}{2} \exp{-kt} \left(1+\exp{-\alpha t} \right)
    + \frac{w}{2}\left(
    \frac{1}{k-\alpha}\left(\exp{-\alpha t}
    -\exp{-kt}
    \right)
+    \frac{1}{k+\alpha}\left(1-\exp{-(k+\alpha)t}
    \right)
    \right)
    \,,
    \label{xRt} \\
\fl
     \ave{x_\phi^2(t)} =    \frac{1}{2}
    \left[
    \int_{-\infty}^\infty \dint{x} \He{0}{\frac{x}{L}}u_0(x)\right.
    \left(
\tikz[baseline=-2.5pt,scale=0.75]{
\draw[Aactivity] (0.75,0) -- (-0.75,0) node [at start, above] {$_0,0$} node [at end, above] {$_0,t$};
}+
\tikz[baseline=-2.5pt,scale=0.75]{
\draw[substrate] (0.75,0) -- (-0.75,0) node [at start, above] {$_0,0$} node [at end, above] {$_0,t$};
}\right)
\tildeu_0(x_0) \nonumber\\ 
+\int_{-\infty}^\infty \dint{x} \He{2}{\frac{x}{L}}u_2(x)
\left(\left(
\tikz[baseline=-2.5pt,scale=0.75]{
\draw[Aactivity] (0.75,0) -- (-0.75,0) node [at start, above] {$_2,0$} node [at end, above] {$_2,t$};
}+
\tikz[baseline=-2.5pt,scale=0.75]{
\draw[substrate] (0.75,0) -- (-0.75,0) node [at start, above] {$_2,0$} node [at end, above] {$_2,t$};
}\right)\tildeu_2(x_0)\right.\nonumber\\ 
+ \left(
\tikz[baseline=-2.5pt,scale=0.75]{
\draw[Aactivity] (0,0) -- (-0.75,0) node [at end, above] {$_2,t$};
\draw[substrate] (0.75,0) -- (0,0) node [at start, above] {$_1,0$};
}
+
\tikz[baseline=-2.5pt,scale=0.75]{
\draw[substrate] (0,0) -- (-0.75,0) node [at end, above] {$_2,t$};
\draw[Aactivity] (0.75,0) -- (0,0) node [at start, above] {$_1,0$};
}
\right)\tildeu_1(x_0)
+\left.\left.
\left(
\tikz[baseline=-2.5pt,scale=0.75]{
\draw[Aactivity] (-0.3,0) -- (-0.75,0) node [at end, above] {$_2,t$};
\draw[substrate] (0.3,0) -- (-0.3,0);
\draw[Aactivity] (0.75,0) -- (0.3,0) node [at start, above] {$_0,0$};
} +
\tikz[baseline=-2.5pt,scale=0.75]{
\draw[substrate] (-0.15,0) -- (-0.75,0) node [at end, above] {$_2,t$};
\draw[Aactivity] (0.15,0) -- (-0.15,0);
\draw[substrate] (0.75,0) -- (0.15,0) node [at start, above] {$_0,0$};
}
\right)\tildeu_0(x_0)\right)\right]\,.
\label{x2monstruosity}
\end{eqnarray}
\end{numparts}
From the propagators in \somepropagatorsclosedform,
we see that at stationarity, only those terms remain where 
the Doi-shifted creation field is 
$\tilderho_0$. Then, \Esref{all_moments} simplify to
\begin{numparts}
\label{all_moments_asymptotics}
\begin{eqnarray}
    \lim_{t\to\infty} \ave{x_\phi^0(t)} &=& \frac{1}{2}\\
    \lim_{t\to\infty} \ave{x_\phi^1(t)} &=& \frac{1}{2} \frac{w}{k+\alpha}\\
    \lim_{t\to\infty} \ave{x_\phi^2(t)} &=& \frac{D}{2k} +\frac{w^2}{2k(k+\alpha)}\,
\end{eqnarray}
\end{numparts}
at stationarity.

\section{Entropy production} 
\label{sec_entropy}
In this section we derive the internal entropy production
$\entropyProduction$ at stationarity \cite{CocconiETAL:2020,VandenBroeck:2010,MaesETAL:2000,SpinneyFord:2012}.
Other observables we calculate
are in the appendix: 
mean square displacement (\ref{sec:app_MSD}),
two-point correlation function (\ref{sec:app_corr_point}),
two-time correlation function (\ref{sec:app_corr}),
expected velocity (\ref{sec:app_vel}) and
stationary distribution (\ref{sec:app_distribution}).

The internal entropy production 
is defined as \cite{Gaspard:2004, CocconiETAL:2020, Garcia-MillanPruessner:2020}
\begin{equation}
\entropyProduction (t)= \lim_{\tau\to0} \frac{1}{\tau} \int  \dint{x}\dint{y} 
P(x,t)W(x\to y;\tau) 
\log\left(
\frac{P(x,t) W(x\to y;\tau)}{P(y,t) W(y\to x;\tau)}
\right) \,,
\label{entrprod_def}
\end{equation}
where $P(x,t)$ is the probability that the system is 
in state $x$ at time $t$ and
$W(x\to y;\tau)$ is the transition probability of the system to change from state 
$x$ to $y$ in an interval of time 
$\tau$. 
The internal entropy production rate $\entropyProduction$
is non-negative, and it is zero if and only if the
detailed balance condition
$P(x)W(x\to y)=P(y)W(y\to x)$ is satisfied for any
two pairs of states $x$ and $y$.
A positive entropy production rate is thus the signature
of non-equilibrium and it indicates the breakdown of 
time-reversal symmetry.
The entropy production $\entropyProduction(t)$
of a drift-diffusive particle in free space with 
velocity $w$ and diffusion constant $D$ is known
to be $\entropyProduction(t)=1/(2t)+w^2/D$ 
\cite{CocconiETAL:2020}, where the first contribution 
is due to the relaxation to steady state and is
independent of the system parameters, and the second
contribution is due to the steady-state probability
current.

We can anticipate that the stationary entropy production 
rate $\entropyProduction$ of an RnT particle in a
harmonic potential is positive given that, between
tumbles, the particle is drift-diffusive and, therefore,
there is locally a perpetual current. Moreover, a
particle's forward trajectory, such as in \fref{sketchRnT},
is distinct from its backwards trajectory, which 
indicates the breakdown of time-reversal symmetry.

Given the RnT particle is confined in a potential, its probability distribution develops into a stationary state,
see \ref{sec:app_distribution}. We denote the stationary distribution by
$P(x)=\lim_{t\to\infty}P(x,t)$. In the limit 
$t\to\infty$, the following identity holds for a Markov
process
\begin{equation}
\int\dint{x} W(x\to y;\tau)P(x) = P(y) \,.
\label{identity_entprod1}
\end{equation}
Using \eref{identity_entprod1}, the Markovian property
\begin{equation}
\int\dint{x} W(y\to x;\tau) = 1 \,,
\label{identity_entprod2}
\end{equation}
and the convention
${W}(y\to x;0) = \delta(x-y)$, 
\Eref{entrprod_def} at stationarity simplifies to

\begin{equation}
\lim_{t\to\infty}\entropyProduction(t) =  \lim_{\tau\to0}
\int  \dint{x}\dint{y} P(x)
\dot{W}(x\to y;\tau) 
\log\left(
\frac{W(x\to y;\tau)}{W(y\to x;\tau) }
\right) \,.
\label{entprod_noP}
\end{equation}

Since the entropy production involves the limit 
$\tau\to0$ of the transition rate $\dot W=\frac{\plaind}{\plaind \tau} W$,
the entropy production crucially draws on the 
microscopic dynamics of the process. We can therefore
focus on lower order contributions to $\dot W$ in 
$\tau$ and neglect higher order contributions.

For an RnT particle, all possible transitions 
between states involve a displacement (run) and/or
a change in the direction of the drift (tumble).
Given that, at stationarity, there is a symmetry
between right- and left-moving particles, we can
summarise the contributions to the entropy production
\eref{entprod_noP} as
\begin{eqnarray}
\fl
  \lim_{t\to\infty}   \entropyProduction(t) =   
\lim_{\tau\to0}
2 \int  \dint{x}\dint{y} 
 P_\phi(x) \left[
\dot{W}(x\to y,\phi;\tau) 
\log\left(
\frac{W(x\to y,\phi;\tau)}{W(y\to x,\phi;\tau) }
\right) \right. \nonumber\\
\left.
+ 
\dot{W}(x\to y,\phi\to\psi;\tau) 
\log\left(
\frac{W(x\to y,\phi\to\psi;\tau)}{W(y\to x,\psi\to\phi;\tau) }
\right)\right]\,,
\label{dotSi_all}
\end{eqnarray}
where $P_\phi(x)=\lim_{t\to\infty}\save{\phi(x,t)\tildephi(x_0,0)}$.
The first term in the square bracket in \eref{dotSi_all} corresponds to the displacement of a right-moving particle from $x$ to $y$
with transition probability
$W(x\to y,\phi;\tau)=\save{\phi(y,\tau)\tildephi(x,0)}$.
The second term in the square bracket of \eref{dotSi_all} 
corresponds to the displacement of a particle
from $x$ to $y$ that starts as right-moving and
ends as left-moving, with transition probability 
$W(x\to y,\phi\to\psi;\tau)=\save{\psi(y,\tau)\tildephi(x,0)}$.
These two transitions include any intermediate state where there may
be displacement and transmutation,
although their transition probabilities are of
higher order in $\tau$ and therefore can be
neglected. We therefore use the short-time propagators in \eref{short_props} \cite{CocconiETAL:2020,Garcia-MillanPruessner:2020}.

In the following, we analyse which terms in \eref{dotSi_all} contribute
to the entropy production rate $\entropyProduction$. We first consider the
transition due to transmutation and displacement, whose short-time propagator is 
\eref{S_shortTimeprop}. By symmetry, we see that
the transition
probability  corresponding to displacement 
and transmutation in the short-time limit is
\begin{equation}
        W(x\to y,\phi\to\psi;\tau)
    = W(y\to x,\psi\to\phi;\tau)
    =    \frac{1}{\sqrt{4\pi D\tau}}
    \exp{-\frac{(y-x)^2}{4D\tau}}
    \left(\frac{\alpha}{2}\tau
    +\mO\left( \tau^2 \right)
    \right) \,,
\end{equation}
which is equal for the forward and backward trajectories. As the 
logarithm vanishes, the second
term in \eref{dotSi_all} does not contribute.

The transition associated to 
the displacement of a right-moving particle from $x$ to $y$
is similar to that of a drift-diffusive
particle in a harmonic potential, so we expect the first term in
\eref{dotSi_all}
to contribute to the entropy production. 
Using the short-time propagator of a right-moving 
particle 
$W(x\to y,\phi;\tau)\simeq\save{\phi(y,\tau)\tildephi(x,0)}_1$
in \eref{R_shortTimeprop_exact},
the first logarithm in \Eref{dotSi_all} is, 
\begin{equation}
\lim_{\tau\to0}\log\left(
\frac{\save{\phi(y,\tau)\tildephi(x,0)}_1}{\save{\phi(x,\tau)\tildephi(y,0)}_1 }
\right) =\frac{y-x}{D} \left( w - k \frac{x+y}{2}\right) \,.
\label{R_log}
\end{equation}
From the short-time propagator in \eref{R_shortTimeprop} and,
equivalently, from the Fokker-Planck equation \eref{FPr}, the kernel
$\dot{W}$
that we need in \eref{dotSi_all} is
\begin{equation}
     \lim_{\tau\to0}\dot{W}(x\to y,\phi;\tau) =
      D\delta''(x-y)  - (w-ky)\delta'(x-y) -\frac{\alpha}{2}\delta(x-y)\,.
      \label{kernel}
\end{equation}

\begin{figure}
    \centering
    \includegraphics[width=\textwidth]{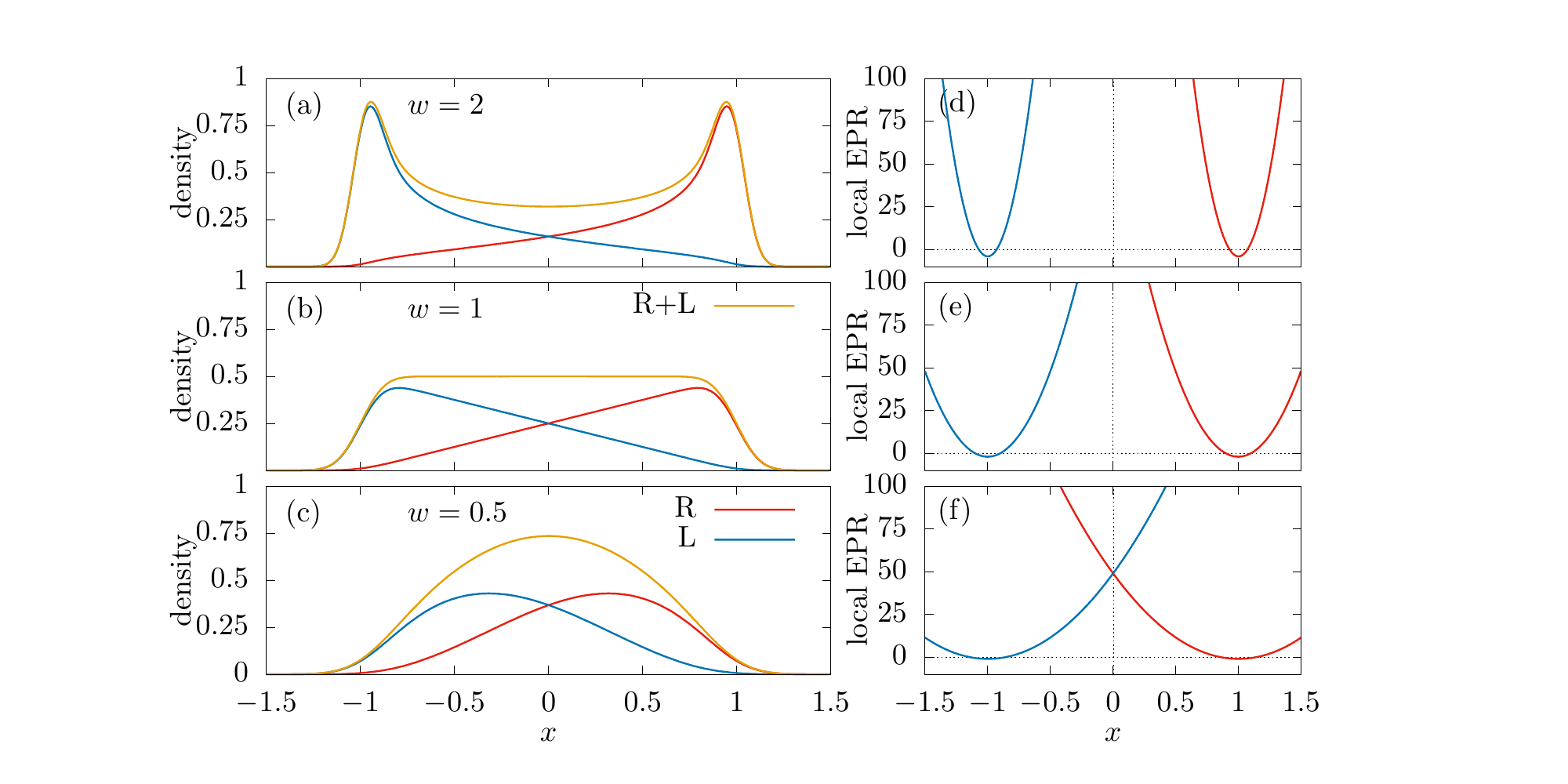}
    \caption{ \textbf{Stationary probability density and local entropy production rate $\sigma(x)$ of a right-moving (R) and left-moving (L) particle},
    \Esref{localEPR}, \eref{stat_distr} and \eref{stat_distr_phi},
    with $D=0.01$, $\alpha=2$.
    In (a) and (d) $w=2$, $k=2$; 
    in (b) and (e) $w=1$, $k=1$; 
    and in (c) and (f) $w=0.5$, $k=0.5$.
    We used multiple-precision floating-point arithmetic
 \cite{FousseETAL:2007} to implement Hermite polynomials
 up to $\He{10^4}{x}$
 based on the GNU Scientific Library implementation
 \cite{GalassiETAL:2009}, see \fref{RnTdistr}.}
    \label{fig_dens_EPR}
\end{figure}

\begin{figure}
    \centering
    \includegraphics[width=0.6\textwidth]{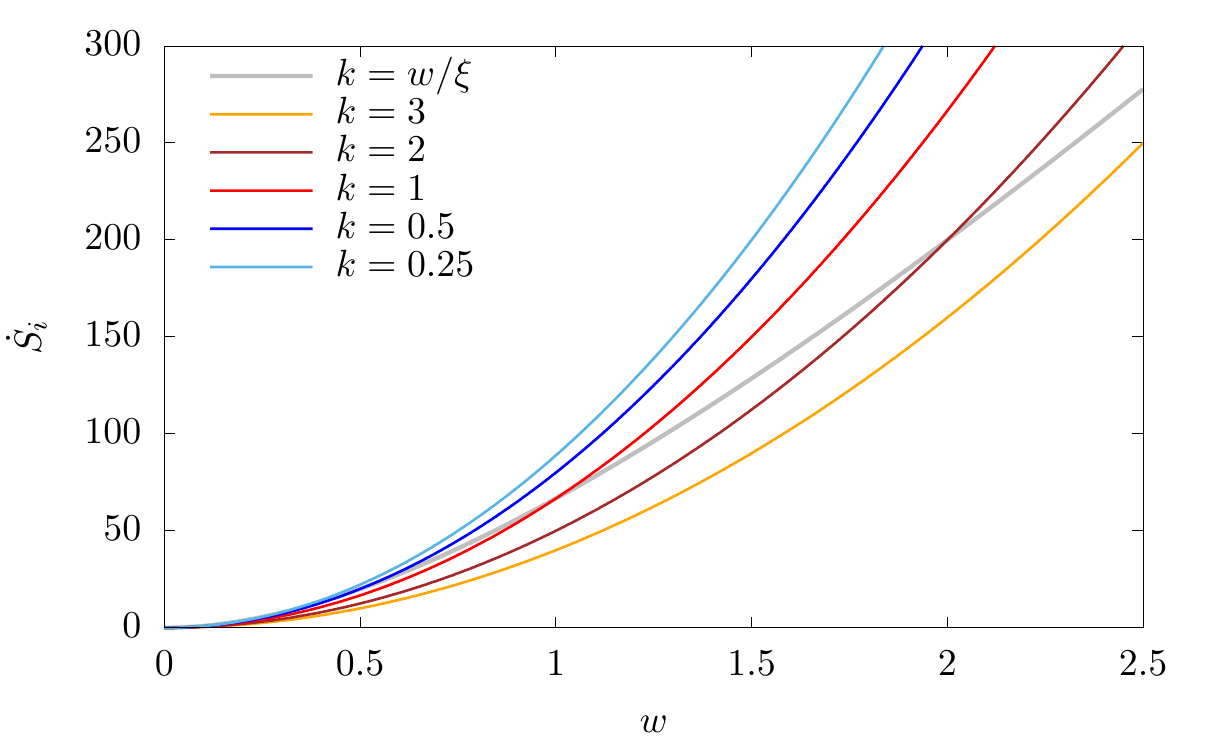}
    \caption{ \textbf{Stationary internal
entropy production rate} of an RnT particle in a harmonic potential
as a function of $w$, \Eref{keyresult},
with $\alpha=2$, $D=0.01$, $\xi=1$ and varying $k$.
The points where the grey line crosses the 
 lines with $k=0.5$, $k=1$ and $k=2$ give
 $\entropyProduction$
in the examples shown in \fref{fig_dens_EPR}.}
    \label{fig_dotSi}
\end{figure}

Using \eref{R_log} and \eref{kernel},  \Eref{dotSi_all} simplifies to
\begin{equation}
\entropyProduction
= 2\int\dint{x} \Pphi(x) \left( \frac{1}{D}(w-kx)^2 -k
\right)
\,,
\label{entProd_moms}
\end{equation}
which shows that the local entropy production rate  \cite{Razin:2020,NardiniETAL:2017}
for right-moving particles is
\begin{equation}
    \sigma_\phi(x) =  \frac{1}{D}(w-kx)^2 -k \,,
\label{localEPR}
\end{equation}
and $\sigma_\psi(x)=\sigma_\phi(-x)$
for left-moving particles, see \fref{fig_dens_EPR}.
The local entropy production rate $\sigma_\phi$ is minimal 
at the characteristic point $\xi=w/k$, where a right-moving particle has a zero expected 
velocity 
because the self-propulsion equals the force exerted
by the potential, 
and its motion is entirely due to the thermal
noise.
We can calculate \eref{entProd_moms}
using the moments derived in 
Sec.~\ref{sec:app_momsposition}.
On the basis of \Esref{all_moments_asymptotics},  the total stationary internal
entropy production 
is 
\begin{equation}
    \entropyProduction =  \frac{ \alpha w^2}{D(k+\alpha)} \,,
    \label{keyresult}
\end{equation}
see \fref{fig_dotSi}.

\section{Discussion and conclusions}
\label{sec_disc}
In this paper we have used the Doi-Peliti framework to describe an RnT particle in a harmonic potential and
calculate its entropy production \eref{keyresult}
and other observables (such as \eqref{short_props}, \eqref{all_moments_asymptotics},
\eqref{msd_stationary},
\eqref{welocity_stationary},
\eqref{twopcorrF},
\eqref{xxcorr}, \eqref{stat_distr}
and
\eqref{stat_distr_phi})
in closed form.
The key result in \Eref{keyresult} shows that the stationary internal
entropy production of an RnT particle is proportional
to that of a drift-diffusive particle in free space, where 
$\entropyProduction = w^2/D$ \cite{CocconiETAL:2020}. 
In the presence of an external harmonic potential ($k>0$),
the entropy production is always smaller than that of a free particle.
If there is no tumbling ($\alpha=0$), then the system is an 
Ornstein-Uhlenbeck process, which is at equilibrium
and therefore produces no entropy.

The positive entropy production rate implies the breaking
of time symmetry whereby forward and backward trajectories 
are distinguishable \cite{CocconiETAL:2020}. This is visible in the  
trajectory of an RnT particle, such as in \fref{sketchRnT}, 
where we can see that the particle runs fast
when "going down" the potential and it slows down  as it
moves up the steep slope of the potential.

The present example of an RnT particle illustrates the power of field theories that
capture the microscopic dynamics to deal with active systems.
Since the large scale behaviour of an RnT particle
is that of a diffusive particle,
by studying an effective theory that captures the large scale
only, we would obtain that the entropy production is zero,
in contradiction with our result \eref{keyresult}.

Finally, deriving the Doi-Peliti field theory  
presents an important technical challenge. Due to the external 
harmonic potential, the action functional is semi-local 
both in real space and in Fourier space. Instead, to diagonalise the action we decomposed
the fields in a basis of Hermite polynomials following the spirit of
the harmonic oscillator \cite{RiskenFrank:1996}.

\ack
We would like to thank Umut \"Ozer for his contribution to an earlier exploration on the field theory of an RnT particle in a harmonic potential. We are grateful to Ziluo Zhang for many enlightening discussions about the field theory of an RnT particle in two dimensions. We would also like to thank Marius Bothe, Benjamin Walter, Zigan Zhen, Luca Cocconi, Fr\'ed\'eric van Wijland, Yuting Irene Li, Patrick Pietzonka and Davide Venturelli for fruitful discussions.
This work was funded in part by the European Research Council under the EU’s Horizon 2020 Programme, grant number 740269.

\appendix

\section{Hermite polynomials} 
\label{sec:sec_Hermite}
The definition of Hermite polynomials \cite{MagnusOberhettingerSoni:1966}
 we use in this paper is
\begin{equation}
\He{n}{x} = (-1)^n \, \exp{\frac{x^2}{2}} \, \frac{\plaind^n}{\plaind x^n}\left(\exp{-\frac{x^2}{2}}\right)\,,
\end{equation}
where $x\in\mathbb{R}$, $n\in\mathbb{N}\union\{0\}$.
Some of the properties
of Hermite polynomials that we  use
are listed in the following 
\cite{MagnusOberhettingerSoni:1966}.

\paragraph{Orthogonality:}
Hermite polynomials are orthogonal with respect to the weight function $f(x) = \exp{-{x^2}/{2}}$, 
\begin{equation}
\int_{-\infty}^\infty\dint{x} \exp{-\frac{x^2}{2}}\He{n}{x}\He{m}{x} = \sqrt{2\pi} \,n!\, \delta_{n,m} \,,
\label{He_orth}
\end{equation}
where $\delta_{n,m}$ is the Kronecker delta. 

\paragraph{Hermite's differential equation:}
Hermite polynomials $\He{n}{x}$ are eigenfunctions of the differential operator 
\begin{equation}
H'' - xH' = -\mu H\,,
\end{equation}
for non-negative integer eigenvalues $\mu_n=n$.
Using this equation, we can show that the
 Hermite functions $u_n(x) = f(x/L)\He{n}{x/L}$ are solution
to  the eigenvalue problem
$\mL[u] = u''+(x\,u)'/L^2=\lambda u$ in Sec.~\ref{sec_rnt01}, \Eref{def_Lop}, with eigenvalues $\lambda_n = - n/L^2$
\cite{RubinPruessnerPavliotis:2014}.

\paragraph{Recurrence relation:}
Hermite polynomials satisfy the recurrence relation
\begin{equation}
    \He{n+1}{x} = 
    x \He{n}{x} - H_n'(x) \,.
    \label{recursive}
\end{equation}

\paragraph{Mehler's formula:}
Hermite polynomials satisfy the following identity
 \cite{MagnusOberhettingerSoni:1966,RiskenFrank:1996},
\begin{equation}
    \exp{-\frac{x^2}{2}}\sum_{n=0}^\infty \frac{s^n}{n!}
    \He{n}{x} \He{n}{y} 
    = \frac{1}{\sqrt{1-s^2}}
    \exp{-\frac{(x-ys)^2}{2\left(1-s^2\right)}}\,.
    \label{Mehler}
\end{equation}

\section{Some propagators in closed form}
\label{sec_propclosedform}
We list the  propagators that we have used to calculate in the 
observables. Using the bare
propagators in \eref{propaabb0_all} and the interaction part
of the action in \eref{mA1rhonu}, we obtain the following
propagators in real time,
\begin{eqnarray}
\fl
 \tikz[baseline=-2.5pt,scale=0.75]{
\draw[Aactivity] (0.75,0) -- (-0.75,0) node [at start, above] {$_m,t'$} node [at end, above] {$_n,t$};
}\,
\hat{=}   
     \ave{\rho_n(t)\tilde{\rho}_m(t')}_0  =  \delta_{n,m} L \exp{- mk (t-t')}\,, 
     \label{prop_aan}\\
\fl
     \tikz[baseline=-2.5pt,scale=0.75]{
\draw[substrate] (0.75,0) -- (-0.75,0) node [at start, above] {$_m,t'$} node [at end, above] {$_n,t$};
}\,
\hat{=}    \ave{\nu_n(t)\tilde{\nu}_m(t')}_0  =  \delta_{n,m} L \exp{-\left(mk + \alpha \right)(t-t')}\,,
\label{prop_bbn}\\
\fl
\tikz[baseline=-2.5pt,scale=0.75]{
\draw[Aactivity] (0,0) -- (-0.75,0) node [at end, above] {$_{n},t$};
\draw[substrate] (0.75,0) -- (0,0) node [at start, above] {$_m,t'$};
}\,
\hat{=}   \ave{\rho_{n}(t)\tilde{\nu}_m(t')(-\mA_1)}_0  
=  
    \delta_{n,m+1}
     \frac{w}{k-\alpha}\exp{-mk(t'-t')}
     \left(\exp{-\alpha(t-t')}-\exp{-k(t-t')}\right)\,,
     \label{prop_abn}\\
\fl
\tikz[baseline=-2.5pt,scale=0.75]{
\draw[substrate] (0,0) -- (-0.75,0) node [at end, above] {$_{n},t$};
\draw[Aactivity] (0.75,0) -- (0,0) node [at start, above] {$_m,t'$};
}\,
\hat{=}   \ave{\nu_{n}(t)\tilde{\rho}_m(t')(-\mA_1)}_0  
=  
    \delta_{n,m+1}
    \frac{w}{k+\alpha}
\exp{-mk(t'-t')}\left(
1-\exp{-(k+\alpha)(t'-t')}\right) \,,
\label{prop_ban}\\
\fl
\tikz[baseline=-2.5pt,scale=0.75]{
\draw[Aactivity] (-0.3,0) -- (-0.75,0) node [at end, above] {$_n,t$};
\draw[substrate] (0.3,0) -- (-0.3,0);
\draw[Aactivity] (0.75,0) -- (0.3,0) node [at start, above] {$_m,t'$};
}\,
\hat{=}  \ave{\rho_n(t)\tilde{\rho}_m(t')(-\mA_1)^2}_0 \nonumber\\ 
 =  
    \delta_{n,m+2}
    \frac{w^2}{L}\exp{-mk(t-t')}
    \left(
     \frac{e^{-2k(t-t')}}{2k(k-\alpha)}
     +\frac{e^{-(k+\alpha)(t-t')}}{(k+\alpha)(\alpha-k)}
     +\frac{1}{2k(k+\alpha)}
     \right) \,,\\
\fl
\tikz[baseline=-2.5pt,scale=0.75]{
\draw[substrate] (-0.15,0) -- (-0.75,0) node [at end, above] {$_n,t$};
\draw[Aactivity] (0.15,0) -- (-0.15,0);
\draw[substrate] (0.75,0) -- (0.15,0) node [at start, above] {$_m,t'$};
}\,
\hat{=}  \ave{\nu_n(t)\tilde{\nu}_m(t')(-\mA_1)^2}_0 \nonumber\\ 
=  
    \delta_{n,m+2}
    \frac{w^2}{L}\exp{-mk(t-t')}
\left(
     \frac{e^{-(2k+\alpha)(t-t')}}{2k(k+\alpha)}
     +\frac{e^{-k(t-t')}}{(k+\alpha)(\alpha-k)}
     +\frac{e^{-\alpha(t-t')}}{2k(k-\alpha)}
     \right) \,,
     \label{prop_babn}
\end{eqnarray}
after letting $r\to0$. These propagators are then used 
to calculate the full propagator in
 real space via
\begin{equation}
    \save{\rho(x,t)\tilderho(x',t')} =
     \frac{1}{L^2}\sum_{n,m,N} 
     u_n(x) \tildeu_m(x') 
    \frac{1}{N!}\ave{\rho_n(t)\tilderho_m(t')(-\mA_1)^N}_0\,,
\end{equation}
where $\ptwiddle{\rho}$ may be replaced by $\ptwiddle{\nu}$.
The stationary distribution is derived from this expression in \ref{sec:app_distribution}.

\section{Mean square displacement} \label{sec:app_MSD}
\begin{figure}
\centering
\begin{subfigure}[c]{.49\textwidth}
	\includegraphics[width=\textwidth]{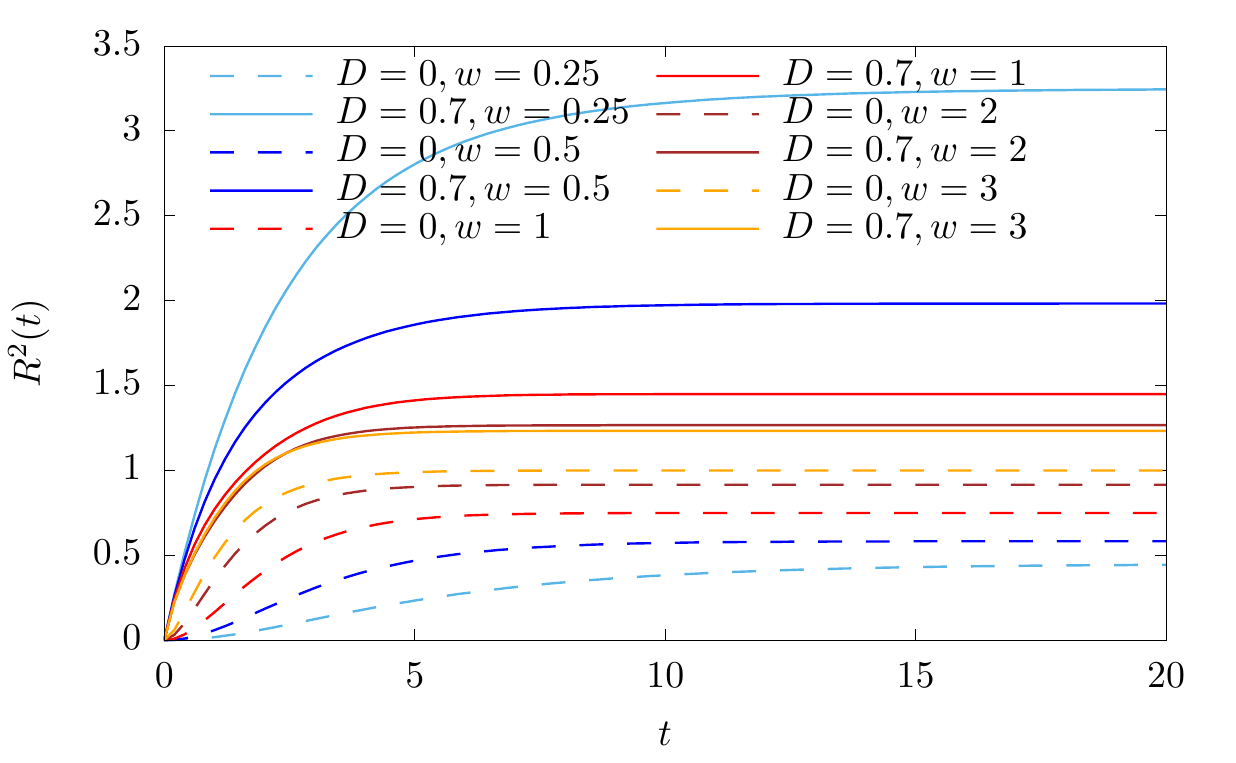} \subcaption{\label{figMSD}}
\end{subfigure}
\begin{subfigure}[c]{.49\textwidth}
	\includegraphics[width=\textwidth]{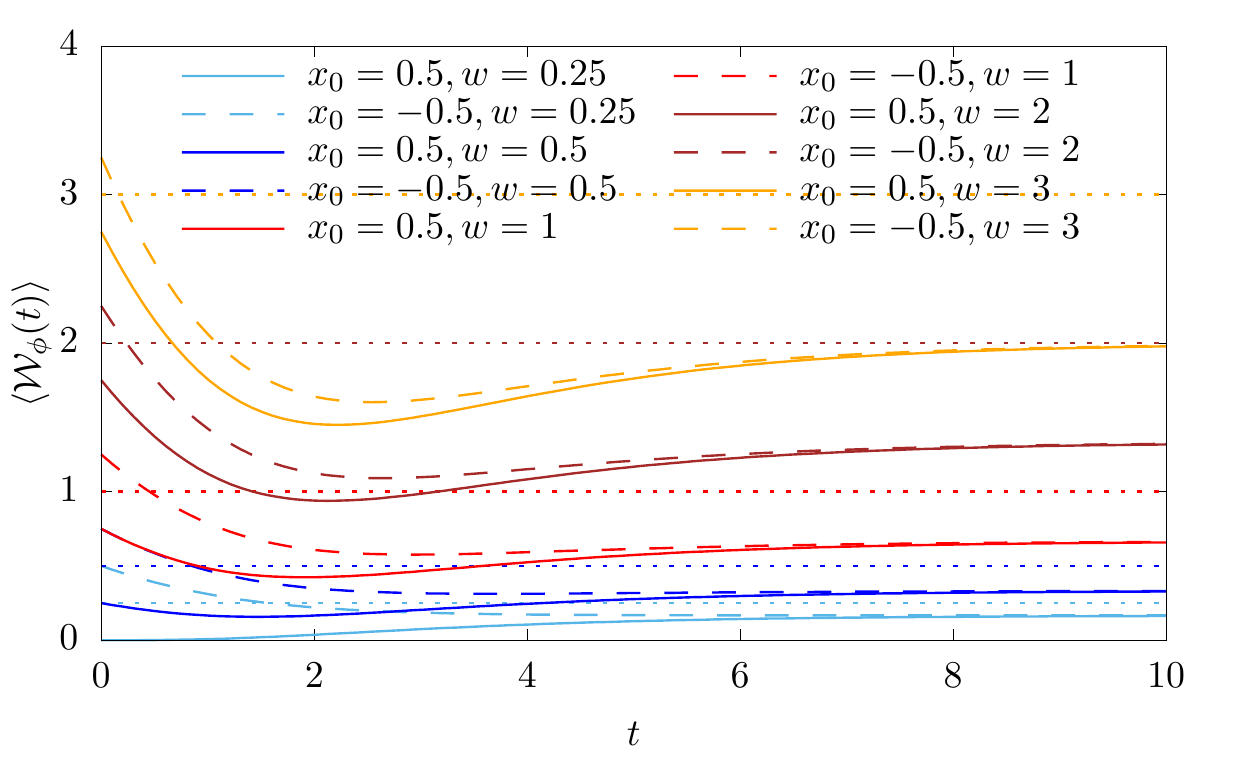} \subcaption{\label{welocity}}
\end{subfigure}
    \caption{\textbf{(\subref{figMSD}) 
    Mean square displacement $R^2(t)$
    } and \textbf{(\subref{welocity}) expected velocity} $\ave{\mW_\phi(t)}$
     of an RnT particle in a harmonic potential.
    In {\bf (\subref{figMSD})}, 
     $x_0=0.5$, $\alpha=1$, $k=w/\xi$
     with $\xi=1$, and a range of $D$,
    using
    \Esref{R2def}, \eref{x_firstmom} and \eref{x_secondmom}.
    In {\bf (\subref{welocity})},
    $k=0.5$, $\alpha=1$ and a range of $x_0$ and $w$,
    using
    \Esref{exp_vel}, \eref{PRt} and \eref{xRt}. The dotted lines show the particle’s self-propulsion speeds $w$, which are the instantaneous velocities at the origin.}
\end{figure}

The mean square displacement is defined
as 
\begin{equation}
    R^2(t) = \ave{\left(x(t)-x_0\right)^2}\,.
    \label{R2def}
\end{equation}
Assuming that the system is initialised
with a right-moving particle, the propagator
that probes for \emph{any} particle at a later
time is
$\bave{\big(\phi(x,t)+\psi(x,t)\big)\tildephi(x_0,0)}$.
Following the same scheme as in 
Sec.~\ref{sec:app_momsposition}, the first and second
moments of the position of an RnT particle are
\begin{eqnarray}
    \ave{x(t)} =  x_0\exp{-kt} 
    + \frac{w}{k-\alpha}\left(\exp{-\alpha t} - \exp{-kt}\right) \,, \label{x_firstmom}\\
    \ave{x^2(t)} = x_0^2\exp{-2kt}+\frac{D}{k}\left(1-\exp{-2kt}\right)
    +2\frac{x_0w}{k-\alpha}\left(\exp{-(k+\alpha)t}-\exp{-2kt}\right) \nonumber\\\qquad
    + 2w^2\left(
     \frac{\exp{-2kt}}{2k(k-\alpha)}
     +\frac{\exp{-(k+\alpha)t}}{(k+\alpha)(\alpha-k)}
     +\frac{1}{2k(k+\alpha)}
     \right) \label{x_secondmom}\,.
\end{eqnarray}
Then, the mean square displacement at stationarity
is
\begin{equation}
\label{msd_stationary}
    \lim_{t\to\infty}R^2(t) = x_0^2 +\frac{D}{k}+ \frac{w^2}{k(k+\alpha)}\,,
\end{equation}
see \fref{figMSD}.

\section{Expected velocity} \label{sec:app_vel}
To calculate the expected velocity of a right-moving
particle, one could na\"ively differentiate the expected position $\ave{x_\phi(t)}$ 
in \Eref{xRt} with respect to time,
\begin{equation}
\partial_t \ave{x_\phi(t)} = \int\dint{x}  x \, \partial_tP_\phi(x,t)\,.
\label{vel_wrong}
\end{equation}
However, this expression fails to capture the 
expected velocity because in the limit 
$t\to\infty$, the stationary distribution
satisfies $\partial_t P_\phi(x,t)=0$, implying that the result
in \eref{vel_wrong} is zero. This is in 
contradiction with the nature of an RnT particle,
which has a perpetual non-zero drift.
Ultimately, the ambiguity in the definition of the velocity is a matter of Ito versus Stratonovich, namely to consider a particle's displacement conditional to its point of departure (Ito), its point of arrival or the average of the two (Stratonovich).

Instead, to calculate the expected velocity of a right-moving particle $\ave{\mW_\phi(t)}$ we 
draw on its local velocity, which is $w-kx$ given its position $x$ prior to its instantaneous departure, as captured by the Fokker-Planck \Eref{rnt2FP}.
The expected velocity
 is then the conditional expectation
\begin{equation}
\ave{\mW_\phi(t)}= \frac{\int \dint{x} (w-kx) \save{\phi(x,t)\tildephi(x_0,t_0)}}{\int \dint{x'} \save{\phi(x',t)\tildephi(x_0,t_0)}} \,,
\label{exp_vel}
\end{equation}
which, using \Esref{PRt} and \eref{xRt}, is, at stationarity
\begin{equation}
\label{welocity_stationary}
\lim_{t\to\infty}\ave{\mW_\phi(t)}= \frac{\alpha w}{k+\alpha} \,,
\end{equation}
see \fref{welocity}.

\section{Two-point correlation function} \label{sec:app_corr_point}
The two-point correlation function $\mF(x,y;t)$
is the observable
\begin{equation}
\label{twopointcorr}
    \mF(x,y;t) = \bave{\big(\phi(x,t)+\psi(x,t)\big)\big(\phi(y,t)+\psi(y,t)\big)\tildephi(x_0,0)} 
    + \bave{\big( \phi(x,t)+\psi(x,t) \big) \tildephi(x_0,0) }
    \delta(x-y)
    \,,
\end{equation}
where, after placing a right-moving 
particle at $x_0$ at time $t_0$, the system is probed for \emph{any} 
particle at positions $x$ and $y$ simultaneously.
The second term contributing only when $x=y$ has its origin in the commutation relation of the creation and annihilation operators \cite{Cardy:2008}.
Since there is exactly
one RnT particle in the system, it cannot be in two 
different positions at the same time and therefore $\mF(x,y;t)=0$
for all $t$ when $x\ne y$. 
Diagrammatically, \eref{twopointcorr} may be written as
\begin{equation}
\mF(x,y;t) \hat{=} \sqrt{2} \left( 
\tikz[baseline=-2.5pt]{
\draw[Aactivity] (-0.2,0.15) -- (-0.75,0.15) ;
\draw[Aactivity] (-0.2,-0.15) -- (-0.75,-0.15) ;
\draw[Aactivity] (0.75,0) -- (0.2,0) ;
\draw[fill=white,draw] (0,0) circle [radius=0.25];
\draw[pattern=north east lines,draw] (0,0) circle [radius=0.25];
}+
\tikz[baseline=-2.5pt]{
\draw[Aactivity] (-0.1,0.15) -- (-0.75,0.15) ;
\draw[Aactivity] (-0.1,-0.15) -- (-0.75,-0.15) ;
\draw[substrate] (0.75,0) -- (0.2,0) ;
\draw[fill=white,draw] (0,0) circle [radius=0.25];
\draw[pattern=north east lines,draw] (0,0) circle [radius=0.25];
}
 \right) 
+ \left(
\tikz[baseline=-2.5pt]{
\draw[Aactivity] (0.75,0) -- (-0.75,0) ;
\draw[black,fill=black] (0,0) circle (0.1);
}+
\tikz[baseline=-2.5pt]{
\draw[Aactivity] (0,0) -- (-0.75,0) ;
\draw[substrate] (0.75,0) -- (0,0) ;
\draw[black,fill=black] (0,0) circle (0.1);
}
\right) \delta(x-y)
 \,,
 \label{twopcorrF}
\end{equation}
which is zero at $x\ne y$ due to a lack of a vertex 
$\tikz[baseline=-2.5pt]{
\draw[Aactivity] (-0.1,0.125) -- (-0.3,0.125) ;
\draw[Aactivity] (-0.1,-0.125) -- (-0.3,-0.125) ;
\draw[Aactivity] (0.3,0) -- (0,0) ;
\draw[fill=white,draw] (0,0) circle [radius=0.2];
\draw[pattern=north east lines,draw] (0,0) circle [radius=0.2];
}$, \ie due to 
the impossibility
of joining a single incoming leg with two out-going legs
because there is no suitable vertex available.
This is an example of how a
Doi-Peliti field theory retains the particle
entity. At $x=y$ the two-point correlation function reduces to the propagators \Esref{propaa_expansion} and \eref{propab_expansion}.

\section{Two-time correlation function} \label{sec:app_corr}
The correlation function $\ave{x(t)x(t')}$, with $t_0< t'< t$, is given by the observable
\begin{equation}
    \ave{x(t)x(t')}
    = \int\dint{x}\dint{x'} x\,x'
    \mG(x,x',x_0;t,t',t_0)\,,
\end{equation}
where the "propagator" is now
  \begin{equation}
     \mG(x,x',x_0;t,t',t_0) =  \big\langle \big[\phi(x,t)+\psi(x,t) \big]  
     \big[\tildephi(x',t')\phi(x',t')+\tildepsi(x',t')\psi(x',t') \big] \tildephi(x_0,t_0)\big\rangle\,.
\end{equation}
This propagator indicates that the system is initialised with a right-moving particle at $x_0$ at time $t_0=0$, and it is let to evolve by an interval of time $t'-t_0$.
At time $t'$, the propagator probes for the presence of a particle at 
$x'$, which involves its annihilation and immediate re-creation.
The system is then let to evolve a further interval of time $t-t'$, at which point the presence
of either species is measured again at position $x$. 
Following the same procedure as in 
Sec.~\ref{sec:app_momsposition},
the two-time correlation function reads
\begin{eqnarray}
\fl
\ave{x(t)x(t')}  
= \,\exp{- k (t-t')} \left[
x_0^2\exp{-2kt'}+\frac{D}{k}\left(1-\exp{-2kt'}\right)
    +2\frac{x_0w}{k-\alpha}\left(\exp{-(k+\alpha)t'}-\exp{-2kt'}\right)
    \right. \nonumber\\\left.
    + 2w^2\left(
     \exp{-2kt'}\frac{1}{2k(k-\alpha)}
     +\exp{-(k+\alpha)t'}\frac{1}{(k+\alpha)(\alpha-k)}
     +\frac{1}{2k(k+\alpha)}
     \right)
\right] \nonumber\\
+ \frac{w}{k+\alpha}\left(
1-\exp{-(k+\alpha)(t-t')}\right)
\left[
\frac{w}{k+\alpha}\left(
1-\exp{-(k+\alpha)t'}\right)
+ x_0 \exp{-(k+\alpha)t'}
\right]\,,
\label{xxcorr}
\end{eqnarray}
see  \cite{Garcia-MillanPhD:2020} 
for details.

\section{Stationary distribution}
\label{sec:app_distribution}
\begin{figure}[t!]
\centering
	\includegraphics[width=0.49\textwidth]{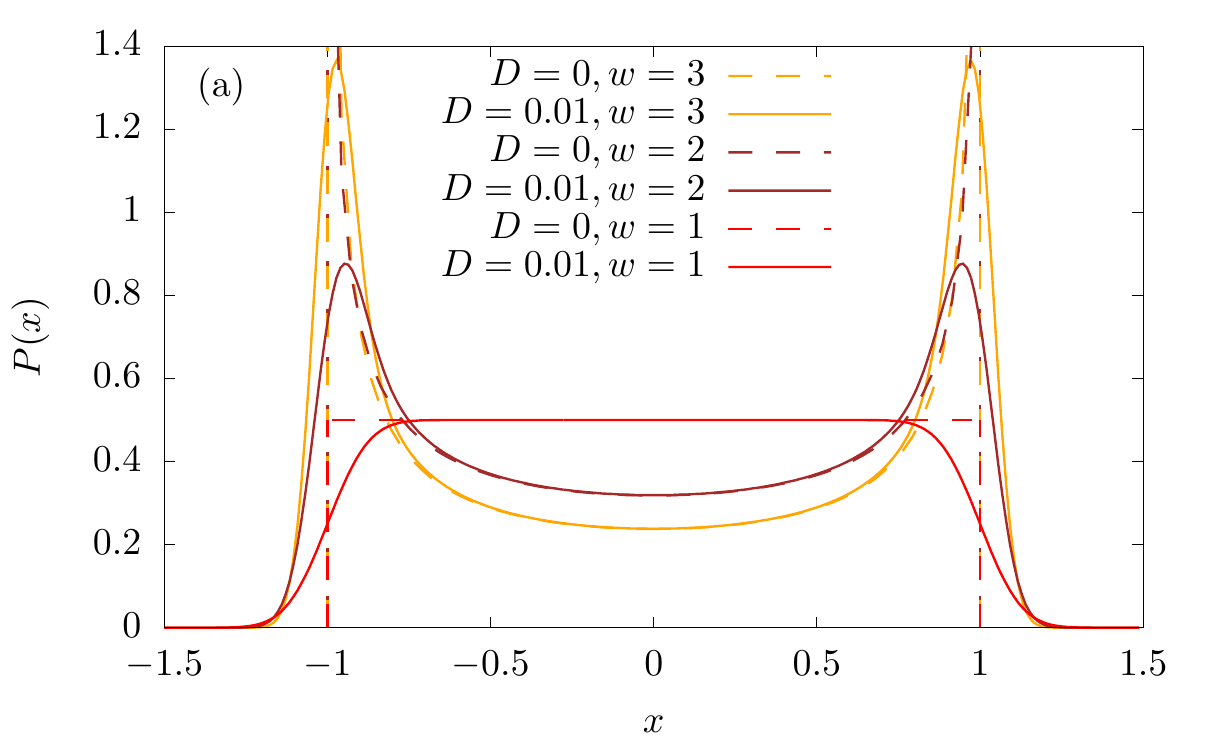} 
	\includegraphics[width=0.49\textwidth]{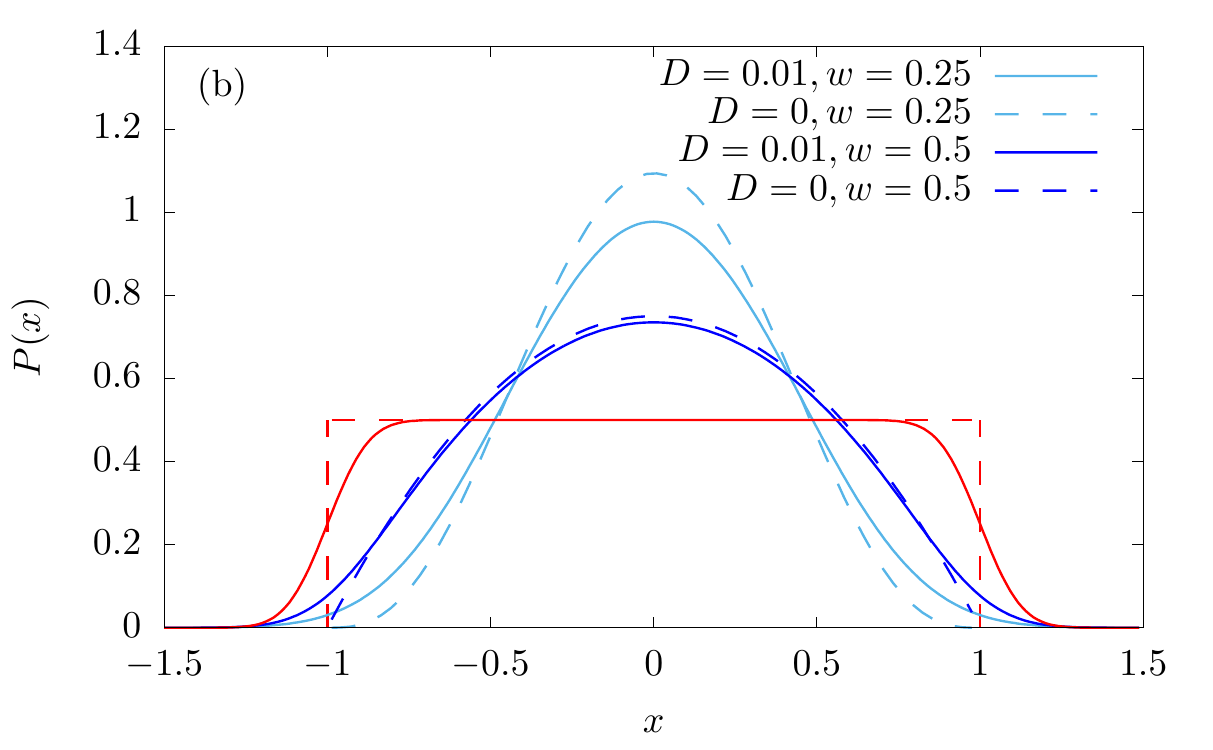} 
\caption{\label{RnTdistr} 
\textbf{Stationary distribution} $P(x)$ of an RnT particle in a harmonic potential
for a range of values of $w$, $k$ and $D$ according to \eref{stat_distr}
for $D>0$ and \eref{stat_distr_D0} for $D=0$.
To ease comparison, we let $\alpha=2$ and $w=\xi k$ so that the limits of the space explored by the 
confined particle in the diffusionless case are $\xi = \pm1$.
In {\bf (a)},
where $k\geq\alpha/2$, the \emph{active} behaviour of the particle is 
manifested  by the  pronounced presence of the particle around 
 $\xi$ for increasing velocity $w$.
In {\bf (b)},
where $k\leq\alpha/2$, the self-propulsion of the particle is less prominent and
its behaviour resembles 
that of a \emph{passive} particle as the velocity $w$ decreases. In fact, for $w=0$, the 
particle is simply a diffusive particle confined in a harmonic potential, which is the
 Ornstein-Uhlenbeck process \cite{UhlenbeckOrnstein:1930}. \index{Ornstein-Uhlenbeck process}
 We used multiple-precision floating-point arithmetic
 \cite{FousseETAL:2007} to implement Hermite polynomials
 up to $\He{10^4}{x}$
 based on the GNU Scientific Library implementation
 \cite{GalassiETAL:2009}, which are needed when the perturbative prefactor $w/\sqrt{kD}$ in \Eref{stat_distr} is large.}
\end{figure}

The distribution of an RnT particle is captured
by the propagator 
$P(x,t) = \save{(\phi(x,t)+\psi(x,t))\tildephi(x_0,0)}$, 
where the system is initialised at $t_0=0$ with a 
right-moving particle at $x_0$
\cite{SinghSabhapanditKundu:2020, DeanMajumdarSchawe:2020}. Diagrammatically,
the particle distribution is
\begin{equation}
    P(x,t) \hat{=}\quad
    \tikz[baseline=-2.5pt]{
\draw[Aactivity] (0.75,0) -- (-0.75,0) ;
\draw[black,fill=black] (0,0) circle (0.1);
}+
\tikz[baseline=-2.5pt]{
\draw[Aactivity] (0,0) -- (-0.75,0) ;
\draw[substrate] (0.75,0) -- (0,0) ;
\draw[black,fill=black] (0,0) circle (0.1);
} \,.
\label{distr_diagr}
\end{equation}
When Fourier transforming back into direct time, all poles $-\imag p$ of all bare propagators of the form $(-\imag \omega + p)^{-1}$, \Eref{propaabb0_all}, eventually feature in the form $\exp{-pt}$.
In the limit $t\to\infty$, from 
\Esref{prop_rhorhoN} and \eref{prop_rhonuN},
we have that any diagram containing $\tildenu$ as
the right, incoming leg 
\tikz[baseline=-2.5pt,scale=0.75]{
\draw[substrate] (0.75,0) -- (0,0) ;
\draw[black,fill=black] (0,0) circle (0.1);
}, decays exponentially in time $t$
(see for instance \eref{prop_bbn}, \eref{prop_abn} and \eref{prop_babn}).
Moreover, the diagrams that have 
\tikz[baseline=-2.5pt,scale=0.75]{
\draw[Aactivity] (0.75,0) -- (0,0) node [at start, above] {$_m,t_0$};
\draw[black,fill=black] (0,0) circle (0.1);
}
as their right, incoming leg decay exponentially with rate $mk$, so only those
with $m=0$ remain in the limit $t\to\infty$. Then, the distribution in
\eref{distr_diagr} reduces to
\begin{equation}
    \lim_{t\to\infty} P(x,t) \hat{=}
    \lim_{t\to\infty} \frac{k}{D}\sum_{n\geq0}
    \tikz[baseline=-2.5pt]{
\draw[Aactivity] (0.75,0) -- (-0.75,0)  node [at start, above] {$_0,t_0$}  node [at end, above] {$_n,t$};
\draw[black,fill=black] (0,0) circle (0.1);
} 
u_n(x)\tildeu_0(x_0)\,,
\label{stat_dens_tot}
\end{equation}
where the sum has contributions only from even $n$ (see 
\Esref{prop_rhorhoN} and \eref{propaa_expansion}). The stationary 
distribution
then reads
\begin{equation}
\lim_{t\to\infty} {P(x,t)} = \sqrt{\frac{k}{{2\pi}D}}\exp{-\frac{kx^2}{2D}}
\left( 1+
 \sum_{\substack{n=2 \\ n\text{ even}}}^\infty \left(\frac{w}{\sqrt{kD}}\right)^{n} \He{n}{\sqrt{\frac{k}{D}}x} \prod_{j=1}^{n} \frac{1}{j+p_j/k}
 \right) \, ,
 \label{stat_distr}
\end{equation}
where $p_j=\alpha$ if $j$ odd and $p_j=r\to0$ otherwise,
see \fref{fig_dens_EPR} and \fref{RnTdistr} \footnote{The stationary  distribution of a non-diffusive
RnT particle follows from the
 coupled Fokker-Planck equations \eref{rnt2FP},
 \begin{equation}
 P(x) = \frac{k \, \Gamma\left(\frac{1}{2}+ \frac{\alpha}{2k} \right)}{\sqrt{\pi} \, w \, \Gamma\left(\frac{\alpha}{2k}\right)} \left[ 1-\left( \frac{k \, x}{w}  \right)^2\right]^{\frac{\alpha}{2k}-1}\,,
\label{stat_distr_D0}
\end{equation}
where $x\in [-w/k,w/k]$ 
\cite{TailleurCates:2008, TailleurCates:2009, Cates:2012, 
BasuETAL:2020, DemaerelMaes:2018, DharETAL:2019}.}.

Similarly, the stationary distribution of a right-moving particle is
\begin{eqnarray}
    \lim_{t\to\infty} P_\phi(x,t) 
  \hat{=}  \lim_{t\to\infty} 
    \frac{1}{2} \Big(
\tikz[baseline=-2.5pt]{
\draw[Aactivity] (0.75,0) -- (-0.75,0) ;
\draw[black,fill=black] (0,0) circle (0.1);
}+
\tikz[baseline=-2.5pt]{
\draw[substrate] (0.75,0) -- (-0.75,0) ;
\draw[black,fill=black] (0,0) circle (0.1);
}+
\tikz[baseline=-2.5pt]{
\draw[Aactivity] (0,0) -- (-0.75,0) ;
\draw[substrate] (0.75,0) -- (0,0) ;
\draw[black,fill=black] (0,0) circle (0.1);
}+\tikz[baseline=-2.5pt]{
\draw[Aactivity] (0.75,0) -- (0,0) ;
\draw[substrate] (0,0) -- (-0.75,0) ;
\draw[black,fill=black] (0,0) circle (0.1);
}\Big)
\nonumber\\
=
\lim_{t\to\infty} 
    \frac{1}{2} \Big(
\tikz[baseline=-2.5pt]{
\draw[Aactivity] (0.75,0) -- (-0.75,0) ;
\draw[black,fill=black] (0,0) circle (0.1);
}+
\tikz[baseline=-2.5pt]{
\draw[Aactivity] (0.75,0) -- (0,0) ;
\draw[substrate] (0,0) -- (-0.75,0) ;
\draw[black,fill=black] (0,0) circle (0.1);
}\Big)\,,
\label{psi_distr_lim}
\end{eqnarray}
whose contribution  $\lim_{t\to\infty} 
\tikz[baseline=-2.5pt,scale=0.5]{
\draw[Aactivity] (0.75,0) -- (-0.75,0) ;
\draw[black,fill=black] (0,0) circle (0.1);
}$
is known from \eref{stat_distr}.
As above, only diagrams that have index $m=0$ remain in the limit $t\to\infty$, so that
\begin{equation}
    \lim_{t\to\infty}  
\tikz[baseline=-2.5pt]{
\draw[Aactivity] (0.75,0) -- (0,0) ;
\draw[substrate] (0,0) -- (-0.75,0) ;
\draw[black,fill=black] (0,0) circle (0.1);
}
=
\lim_{t\to\infty} \frac{k}{D}\sum_{n\geq1}
    \tikz[baseline=-2.5pt]{
    \draw[Aactivity] (0.75,0) -- (0,0) node [at start, above] {$_0,t_0$};
\draw[substrate] (0,0) -- (-0.75,0)  node [at end, above] {$_n,t$};
\draw[black,fill=black] (0,0) circle (0.1);
} 
u_n(x)\tildeu_0(x_0)\,,
\label{sum_nurho_lim}
\end{equation}
which has contributions only from odd $n$, see \Esref{prop_nurhoN} and \eref{propba_expansion}.
\Eref{sum_nurho_lim} has the same form as 
\eref{stat_distr} except that the dummy variable $n$ is odd.
Therefore, the probability distribution in 
\eref{psi_distr_lim} contains the sum over both even and odd
indices $n\geq0$,
\begin{equation}
\lim_{t\to\infty} {P_\phi(x,t)} = \sqrt{\frac{k}{{2\pi}D}}\exp{-\frac{kx^2}{2D}}
\left( 1+
 \sum_{n=1}^\infty \left(\frac{w}{\sqrt{kD}}\right)^{n} \He{n}{\sqrt{\frac{k}{D}}x} \prod_{j=1}^{n} \frac{1}{j+p_j/k}
 \right) \, ,
 \label{stat_distr_phi}
\end{equation}
see \fref{fig_dens_EPR}.

\bibliography{articles,books}
\end{document}